\documentclass[onecollarge]{svjour2}       
\usepackage{graphicx,amsmath,psfrag}

\begin{document}

\title{Strong and weak chaos in weakly nonintegrable\\ 
many-body Hamiltonian systems}

\author{M.~Mulansky \and K.~Ahnert \and A.~Pikovsky and D.~L.~Shepelyansky}
\institute{M.~Mulansky, K.~Ahnert and A.~Pikovsky  
\at Department of Physics and Astronomy, Potsdam University,
  Karl-Liebknecht-Str 24, D-14476, Potsdam-Golm, Germany
\and
D.~L.~Shepelyansky \at Laboratoire de Physique Th\'eorique du CNRS (IRSAMC), 
Universit\'e de Toulouse, UPS, F-31062 Toulouse, France}

\date{March 14, 2011}


\maketitle

\begin{abstract}
We study properties of chaos in generic one-dimensional
nonlinear Hamiltonian lattices comprised  of weakly coupled
 nonlinear oscillators, by numerical simulations
of continuous-time systems and symplectic maps. 
For small coupling, the measure of chaos
is found to be proportional to the coupling strength
and lattice length, with the typical maximal Lyapunov exponent 
being proportional to the square root of coupling.
This strong chaos appears as a result of triplet
resonances between nearby modes. In addition 
to strong chaos we observe  a weakly chaotic component
having much smaller Lyapunov exponent,
the measure of which drops approximately
as a square of the coupling strength 
down to smallest couplings we were able to reach.
We argue that this weak chaos 
is linked  to the regime of fast Arnold diffusion
discussed by Chirikov and Vecheslavov.
In disordered lattices of large size we find a subdiffusive
spreading of initially localized wave packets 
over larger and larger number of modes. 
The relations
between the exponent of this spreading and the 
exponent in the dependence of
the fast Arnold diffusion on coupling strength
are analyzed. We also trace parallels between 
the slow spreading of chaos and deterministic rheology.
 
\end{abstract}

\section{Introduction}
Even 120 years after the fundamental work of 
Poincar\'e \cite{Poincare-1890} and  numerous efforts done after it,
an interplay between order and chaos in high-dimensional Hamiltonian systems
remains a challenging problem. For Hamiltonian dynamics with a few degrees 
of freedom, a clear picture of a separation between chaotic and regular 
(quasiperiodic) regions in
 the phase space~\cite{Chirikov-79,Lichtenberg-Lieberman-92} 
has been confirmed in numerous studies. Much less is known on 
this separation and the structural properties of chaos 
when the number of degrees of freedom becomes large. 
Especially the generic case  of a weak nonlinear coupling of 
initially nonlinear but integrable  degrees of freedom 
remains poorly understood. We will call such systems to be
weakly nonintegrable. Their properties are very
nontrivial since a decrease in nonlinearity/nonintegrability might 
be compensated by an increase of dimensionality of the phase space.

The Kolmogorov-Arnold-Moser (KAM) theory guarantees the existence
of invariant tori at a sufficiently weak nonlinear perturbation
(see e.g. \cite{Chirikov-79,Lichtenberg-Lieberman-92}). 
However, in conservative systems with more than two degrees of freedom
($L>2$) such tori are not isolating and chaos can spreads 
along tiny chaotic layers as it was pointed by Arnold
\cite{Arnold-64}. The mechanism of such a chaotic spreading 
is known under the name of Arnold diffusion 
as coined by Chirikov in 1969
\cite{Chirikov-79,Chirikov-69}. For $L=3$
the rate of Arnold diffusion drops exponentially
with the dimensionless strength of nonlinear coupling $\beta$
\cite{Chirikov-79,Lichtenberg-Lieberman-92,Chirikov-69}.
This is in a qualitative agreement with
a number of  mathematical results
which give rigorous bounds on the spreading rate 
in the limit of asymptotically small  $\beta$
at fixed $L>2$ \cite{Nekhoroshev-77,Lochak-92}.
The mathematical studies of the Arnold diffusion
properties are  actively continued at present (see e.g.
\cite{Kaloshin-08} and Refs. therein). While the mathematical 
results indicate the exponentially small 
rate of Arnold diffusion $D_A$ in the limit 
of small nonlinearity $\beta$ at fixed $L$,
it remains not clear at what realistic values
of nonlinearity such an exponential behavior effectively appears.
The striking results of Chirikov and Vecheslavov,
established by extensive numerical simulations for $4 \leq L \leq 15$
and supported by heuristic arguments
\cite{Chirikov-Vecheslavov-90,Chirikov-Vecheslavov-93,Chirikov-Vecheslavov-97},
show only an algebraic decay of $D_A$ with $\beta$
up to extremely small values of Arnold diffusion coefficient
$D_A \sim 10^{-50}$.
This regime was named by them the fast Arnold diffusion.
These studies have been restricted by $L \leq 15$
and it remains unclear what can happen 
with such a behavior in the limit of larger $L$ with small but fixed $\beta$. 

The question about the properties of Hamiltonian systems 
at large values of $L$ is linked to the 
fundamental problem of dynamical thermalization
and ergodicity in the thermodynamic limit.  
As typical models with a large number of degrees of freedom one considers 
Hamiltonian lattices (or Hamiltonian partial differential equations, 
which, however, live in an infinite-dimensional phase space). 
A striking example of nontrivial
dynamics in weakly nonlinear lattices gives 
the Fermi-Pasta-Ulam problem \cite{Fermi-55}, 
which is still far from being completely resolved despite 
of numerous efforts in its 50-year history 
(see  \cite{Chaos-fpu-05,Gallavotti-08} for a stand around 2004 
and \cite{Benettin-Livi-Ponno-09} for recent advances). 
Moreover, because the FPU model has a special peculiarity 
as being close to an integrable Toda lattice, 
its properties appear to be rather special.
Quite recently, a lot of attention attracted 
disordered nonlinear lattices~\cite{Shepelyansky-93,%
Molina-98,%
Pikovsky-Shepelyansky-08,Garcia-Mata-Shepelyansky-09,%
Flach-Krimer-Skokos-09,Skokos_etal-09,%
Mulansky-Ahnert-Pikovsky-Shepelyansky-09,%
Skokos-Flach-10,Flach-10,Laptyeva-etal-10,%
Mulansky-Pikovsky-10,Johansson-Kopidakis-Aubry-10}
studied in the context of the problem of nonlinear destruction 
of Anderson localization. Here one tries to relate 
the properties of chaos and regularity at small nonlinearities 
to the properties of the spreading of a wave packet over 
the lattice~\cite{Basko-10,Krimer-Flach-10,Pikovsky-Fishman-10}.
Certain mathematical bounds on the rate of spreading
have been obtained \cite{Wang-08,Bourgain-08} 
by the methods similar to those of
Nekhoroshev \cite{Nekhoroshev-77} but
they are available only in the limit of very small
nonlinearity being very far from the regimes
studied in numerical simulations.
In addition, these weakly nonlinear lattices are not generic objects from 
the point of view of weak nonintegrability and the KAM theory, 
since in the limit of small coupling they are reduced 
to a set of linear modes, 
i.e. to a linear quasiperiodic state demonstrating qusiperiodicity 
and pure point spectrum typical of the Anderson localization, 
and not to the generic case with a set of uncoupled 
\textit{nonlinear} modes. 
We note, that in context of the KAM theory, a small perturbation of 
the latter integrable nonlinear system is studied.

In this paper we study properties of a lattice of weakly 
coupled nonlinear oscillators at small coupling and 
large number of degrees of freedom. In the limit of small coupling 
this model reduces to an integrable although strongly nonlinear one, 
demonstrating typically quasiperiodic dynamics.
A nice model of such a setup has been suggested 
by Kaneko and Konishi~\cite{Kaneko-Konishi-89,Konishi-Kaneko-90}, 
it gives a generalization of the Chirikov standard map~\cite{Chirikov-79} 
to a lattice of coupled symplectic maps. 
This model is computationally efficient and allows one a 
rather good numerical characterization of properties 
of regular and chaotic dynamics. Nevertheless, even for this model 
the quantitative properties
are not well-established despite of various 
efforts~\cite{Falcioni-Paladin-Vulpiani-89,Falcioni-91,%
Chirikov-Vecheslavov-93,Chirikov-Vecheslavov-97,Lichtenberg-Aswani-98}. 
Additionally, we study here two models of coupled nonlinear 
continuous-time oscillator lattice where the spreading 
over the lattice can be analyzed at fixed energy. 
Our main conclusions are valid for all these systems.

The plan of the paper is as follows. We start by formulating basic models 
we study in Section~\ref{sec:bm}.
Then in Section~\ref{sec:le} we discuss the properties of 
the largest Lyapunov exponent, especially
the scaling relations in dependence on coupling strength
and system length. In Section~\ref{sec:res} we argue that chaos 
is mainly due to occasional resonances between triples of 
three neighboring oscillators. 
In Section~\ref{sec:md} we discuss statistical properties of chaos, 
focusing on the scaling of the diffusion constant. 
In Section~\ref{sec:spread} we relate this properties to that 
of spreading of a wave packet in an unbounded lattice. 
Finally, in Section~\ref{sec:rheology} a very slow evolution is compared 
to similar effects in the context of rheology.

\section{Basic models}
\label{sec:bm}

Here we introduce three generic models of nonlinear oscillators 
locally coupled in space.
Model A, introduced by Kaneko and 
Konishi~\cite{Kaneko-Konishi-89,Konishi-Kaneko-90},
is a model of coupled symplectic maps
\begin{equation}
\begin{aligned}
\bar{p}_k&=p_k+K[\sin(x_{k+1}-x_k)+\sin(x_{k-1}-x_k)]\;,\quad k=1,\ldots,L\\
\bar{x}_k&=x_k+\bar{p}_k\;.
\end{aligned}
\label{eq:kk}
\end{equation}
with periodic boundary conditions.
Here $p$ is a ``momentum'' and $x$ is a ``phase'' variable.
In the absence of coupling (i.e. for $K=0$) each oscillator 
has a constant frequency $p_k$ that depends on initial conditions, 
so in the whole lattice generally a
quasiperiodic regime with $L$ frequencies establishes. 
For finite $K$ the oscillators are coupled and chaos is possible.

Model B is a strongly nonlinear continuous-time lattice with Hamiltonian
\begin{equation}
\begin{aligned}
H=\sum_{k=1}^L\frac{p_k^2}{2}+\frac{q_k^4}{4}+\frac{\beta}{2}(q_{k+1}-q_k)^2\;.
\end{aligned}
\label{eq:modb}
\end{equation}
Here we also consider a lattice of length $L$ with periodic boundary 
conditions. The coupling parameter $\beta$ plays the same role as $K$. 
Contrary to model A, model B conserves the total energy. 
We normalize the energy in such a way that $E=L$ 
(i.e. density of energy is one), so that $\beta$ and $L$ are 
the only parameters of this model. 

Very similar to the model B is the model C, 
where the coupling between nonlinear modes is also nonlinear, moreover, 
the power of nonlinearity in coupling is stronger than the local one:
\begin{equation}
\begin{aligned}
H=\sum_{k=1}^L\frac{p_k^2}{2}+\frac{\eta_k q_k^4}{4}+\frac{\gamma}{6}(q_{k+1}-q_k)^6\; ,
\end{aligned}
\label{eq:modc}
\end{equation}
where we consider two cases for coefficients with all
$\eta_k=1$ (C1) and random homogeneously distributed
values $0.5 \leq \eta_k \leq 1.5$ (C2).
While we do not expect large difference between models B and C 
in the described setup, where the density of the energy is fixed, 
the situation changes when the total energy is fixed and 
the length of the lattice is increased. In this limit model B 
will become asymptotically linear (effective $\beta$ increases) 
while model C will become asymptotically less and 
less coupled (effective $\gamma$ decreases). This difference is important 
for the implications of chaos for spreading of 
initially localized wave packets, to be discussed in Section~\ref{sec:spread}. 
The randomness of values of $\eta_k$ (model C2) ensures that there are no
regular waves emanating from the main part of the wave packet
in contrast to the case $\eta_k=1$ (model C1) where such wave radiation is
possible \cite{Ahnert-09,Ahnert-10}.

\section{Lyapunov exponents and their scaling}
\label{sec:le}

\subsection{Lyapunov exponents}
The largest Lyapunov exponent (LE) is a standard measure of chaos and 
is easy to calculate~\cite{Lichtenberg-Lieberman-92,Ott-book-92}. 
We have performed 
a statistical analysis of Lyapunov exponents for models A, B, C 
based on an ensemble of random initial conditions. 
For model A we have chosen $0\leq p_k,x_k<2\pi$ 
as independent uniformly distributed. For model B  
we initialized $q_k=0$  and  $p_k$ normally distributed with zero mean, 
after this the values $p_k$ are rescaled such that the total energy 
of the lattice equals $L$ - the number of lattice sites.
For the model C the initialization is done in a similar way.
We used up to several thousands of initial state realizations to obtain a good
statistics in the computation of measure of chaos $P_{ch}$.

\begin{figure}[!hbt]
\centering
\psfrag{xlabel1}[c][c]{\# of realization}
\psfrag{ylabel1}[c][c]{Lyapunov exponent}
\psfrag{xlabel2}[c][c]{\# of realization}
\psfrag{ylabel2}[c][c]{Lyapunov exponent}
(a)\includegraphics[width=0.4\textwidth]{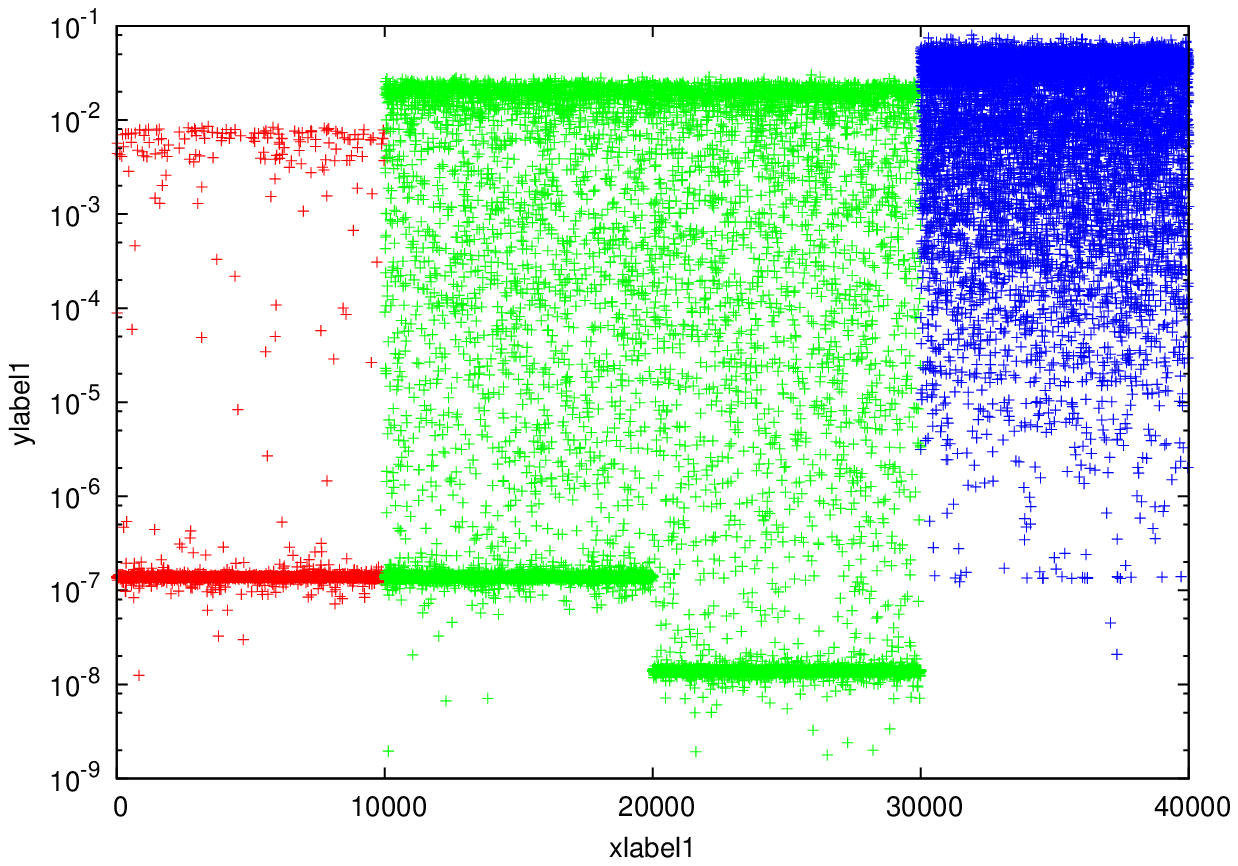}\hfill
(b)\includegraphics[width=0.4\textwidth]{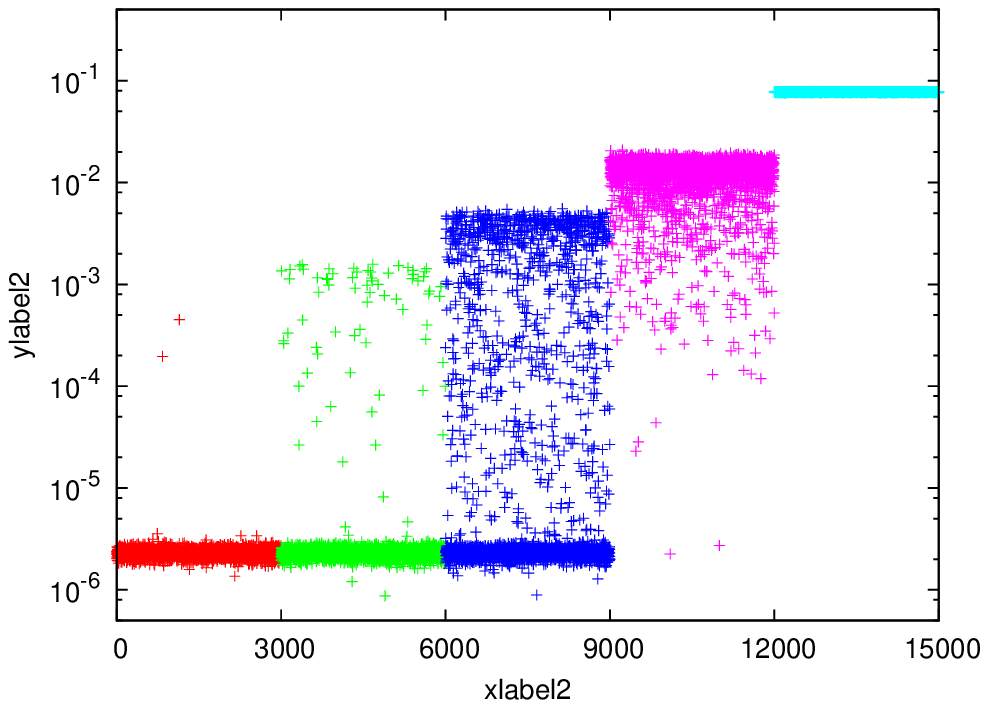}
\caption{(Color online) (a) Calculations of LEs 
for model A with $L=8$. First 10000 points (red): $K=0.001$, 
LE calculated by averaging over time interval $T_{av}=5\cdot10^6$. 
Second 10000 points (green): $K=0.01$, $T_{av}=5\cdot10^6$. 
Next 10000 points (also green): $K=0.01$, $T_{av}=5\cdot10^7$. 
Last 10000 points (blue): $K=0.05$, $T_{av}=5\cdot10^6$.
(b): The same as (a) but for model B with $L=16$, $T_{av}=10^6$ 
and different $\beta$, from left to right: 
$\beta=10^{-5},10^{-4},10^{-3},10^{-2},10^{-1}$, 
for each value of $\beta$ 3000 realizations are shown. }
\label{fig:lere_kk}
\end{figure}

We present the ``raw data'' of these calculations 
for models A and B in Fig.~\ref{fig:lere_kk}. 
Here, for model A in a lattice with $L=8$ one observes 
predominantly chaos for $K=0.05$, 
predominantly regularity for $K=0.001$, and both states depending on 
initial conditions for $K=0.01$. Noteworthy, LE in the case of 
regularity does not vanish but attains very small values, 
with the cutoff appearing due to a finite integration time. 
In the middle part of 
Fig.~\ref{fig:lere_kk}(a) one can see that increasing the integration time by factor 
10 roughly decreases this lower cutoff in the Lyapunov exponent by factor 10. For any fixed $T_{av}$, 
basing on inspection, one easily chooses a threshold in LE that 
separates chaos from regularity. Of course, there are realizations 
with values around these thresholds that cannot be resolved within 
the integration time used, but their statistical relevance is not significant. 
Essentially the same picture is observed for models B (Fig.~\ref{fig:lere_kk}b)
and model C (data not shown).

\subsection{Scaling of probability to observe chaos}
\label{sec:sr}

\begin{figure}[!hbt]
\centering
\psfrag{xlabel0}{$K$}
\psfrag{xlabel1}{$K$}
\psfrag{xlabel2}{$K\cdot L$}
\psfrag{ylabel0}[c][c]{$P_{ch}$}
\psfrag{ylabel1}[c][c]{$P_{ch}$}
\psfrag{ylabel2}[c][c]{$P_{ch}$}
\includegraphics[width=0.3\textwidth]{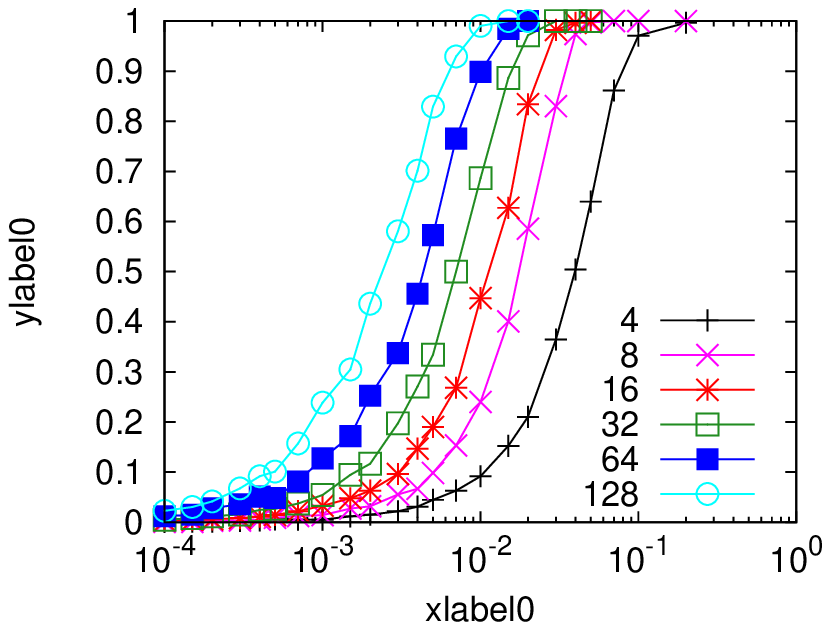}\hfill
\includegraphics[width=0.3\textwidth]{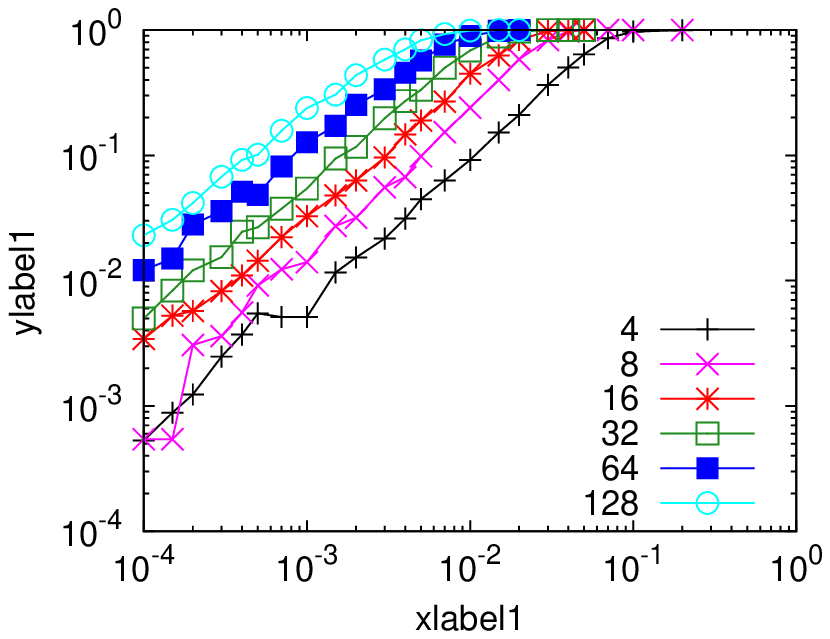}\hfill
\includegraphics[width=0.3\textwidth]{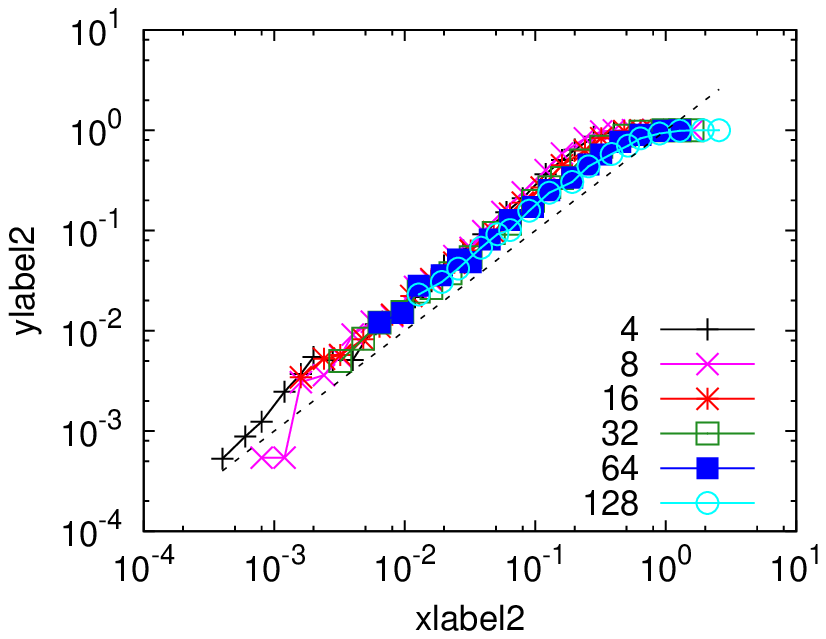}
\caption{Calculations of $P_{ch}$ vs $K$ for model A, 
demonstrating the scaling $P_{ch}\sim KL$ for small $K$. 
The middle panel shows the same data as the left one but 
in a logarithmic scale, while the left panel shows $P_{ch}$ 
as a function of the product $KL$.   
The dashed line on the right panel is $P_{ch}=KL$.}
\label{fig:mc_kk}
\end{figure}

\begin{figure}[!hbt]
\centering
\psfrag{xlabel0}{$\beta$}
\psfrag{xlabel1}{$\beta$}
\psfrag{xlabel2}{$\beta\cdot L$}
\psfrag{ylabel0}[c][c]{$P_{ch}$}
\psfrag{ylabel1}[c][c]{$P_{ch}$}
\psfrag{ylabel2}[c][c]{$P_{ch}$}
\includegraphics[width=0.3\textwidth]{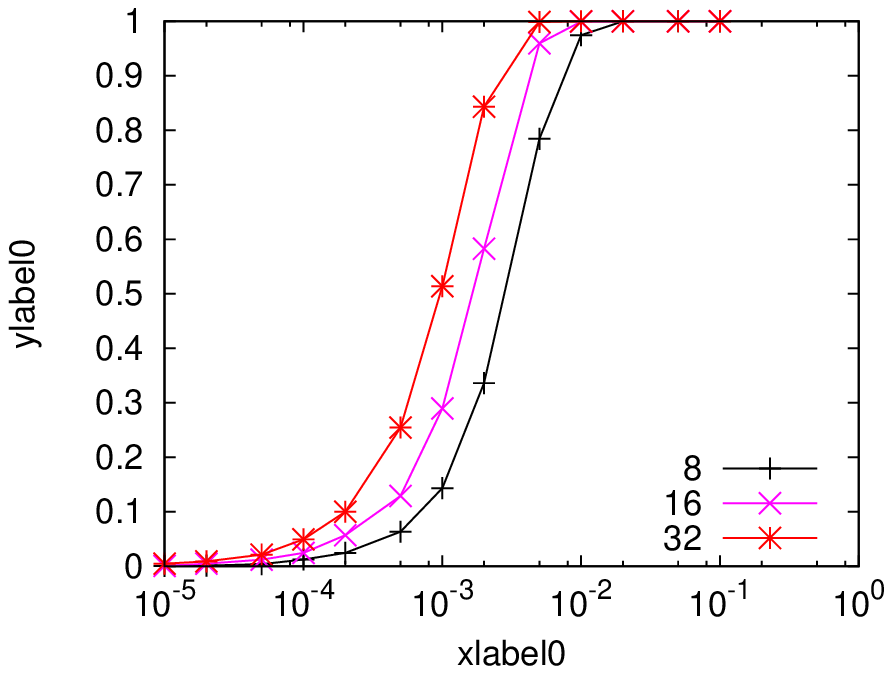}\hfill
\includegraphics[width=0.3\textwidth]{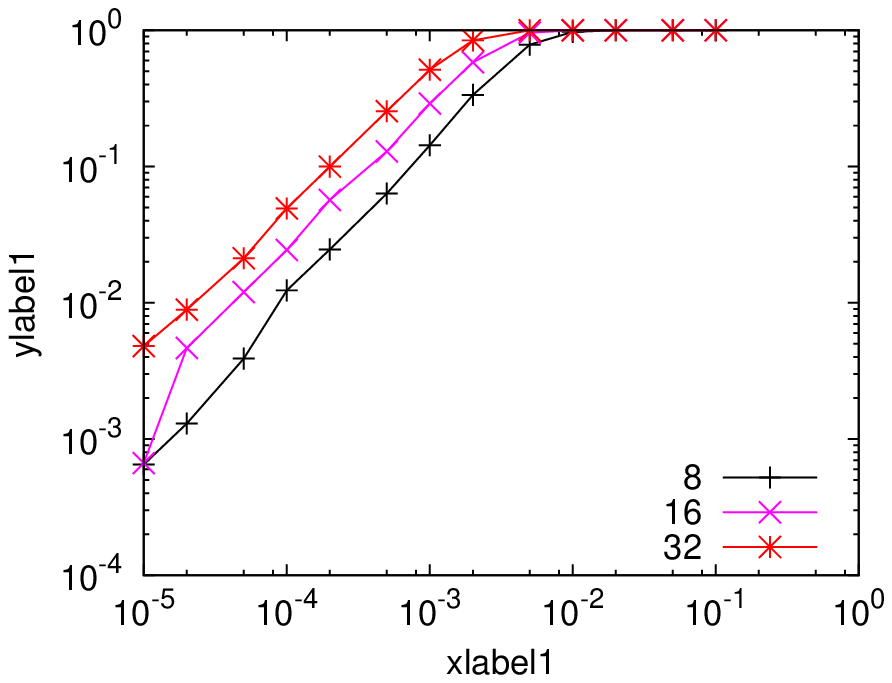}\hfill
\includegraphics[width=0.3\textwidth]{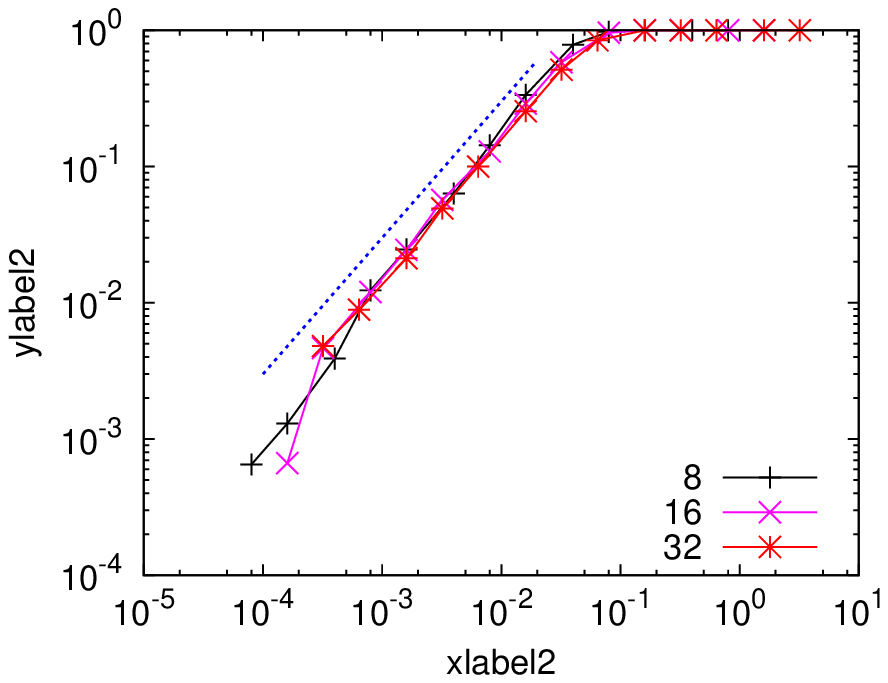}
\caption{The same as Fig.~\ref{fig:mc_kk} but for model B.}
\label{fig:mc_nl}
\end{figure}

\begin{figure}[!hbt]
\centering
\psfrag{xlabel1}{$\gamma$}
\psfrag{xlabel2}{$\gamma$}
\psfrag{xlabel3}{$\gamma\cdot L$}
\psfrag{ylabel1}[c][c]{$P_{ch}$}
\psfrag{ylabel2}[c][c]{$P_{ch}$}
\psfrag{ylabel3}[c][c]{$P_{ch}$}
\includegraphics[height=0.3\textwidth,angle=270]{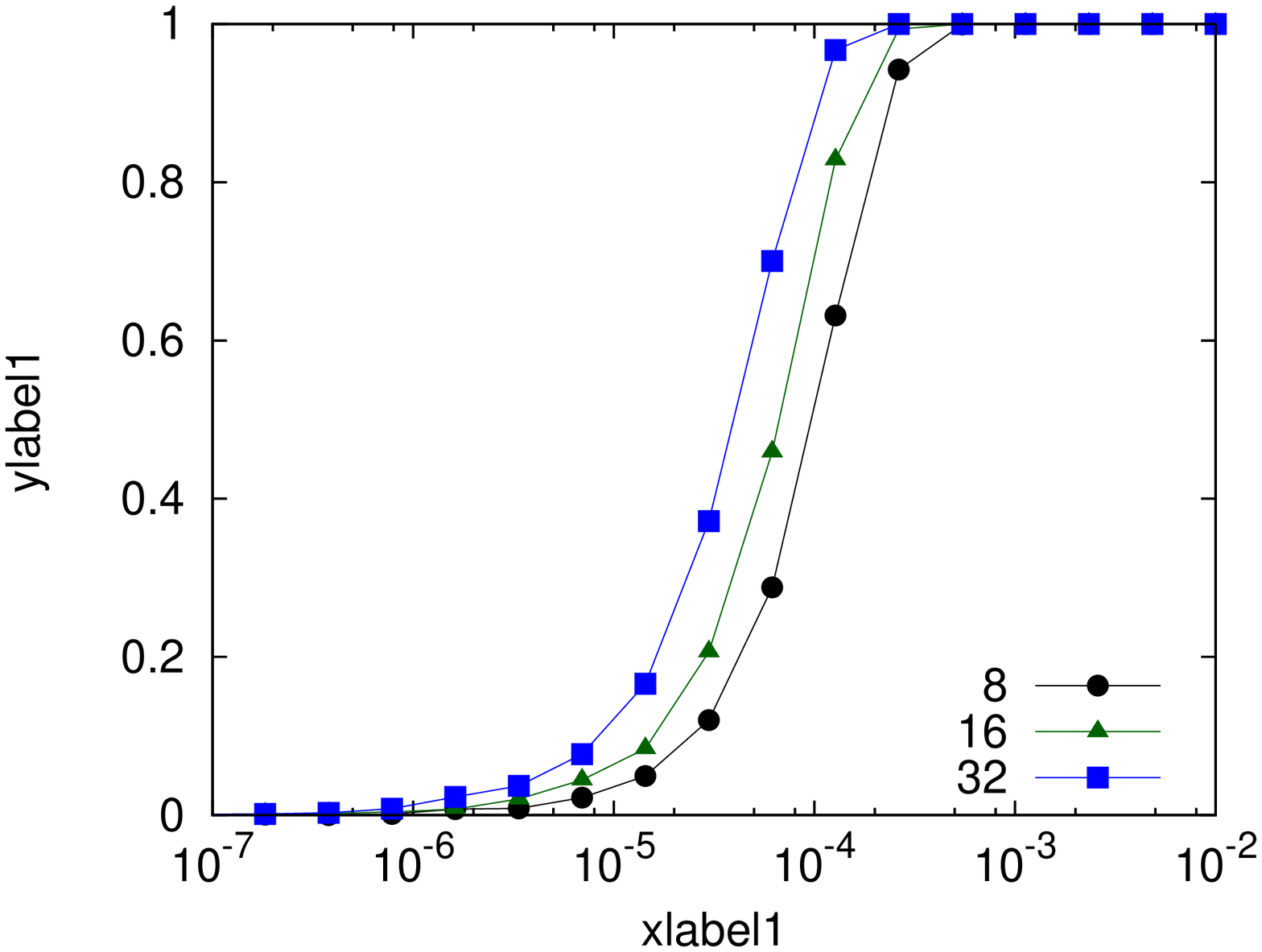}\hfill
\includegraphics[height=0.3\textwidth,angle=270]{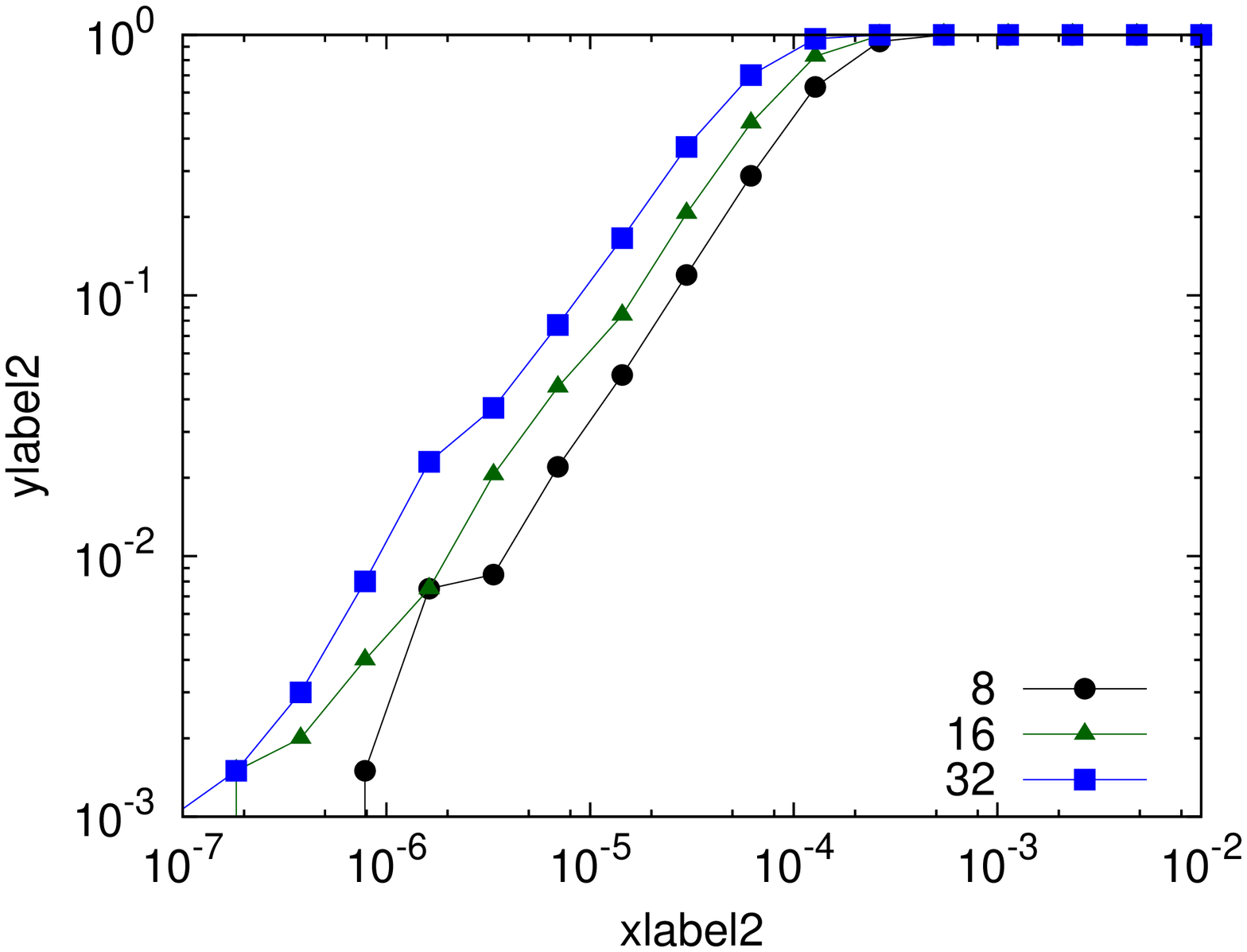}\hfill
\includegraphics[height=0.3\textwidth,angle=270]{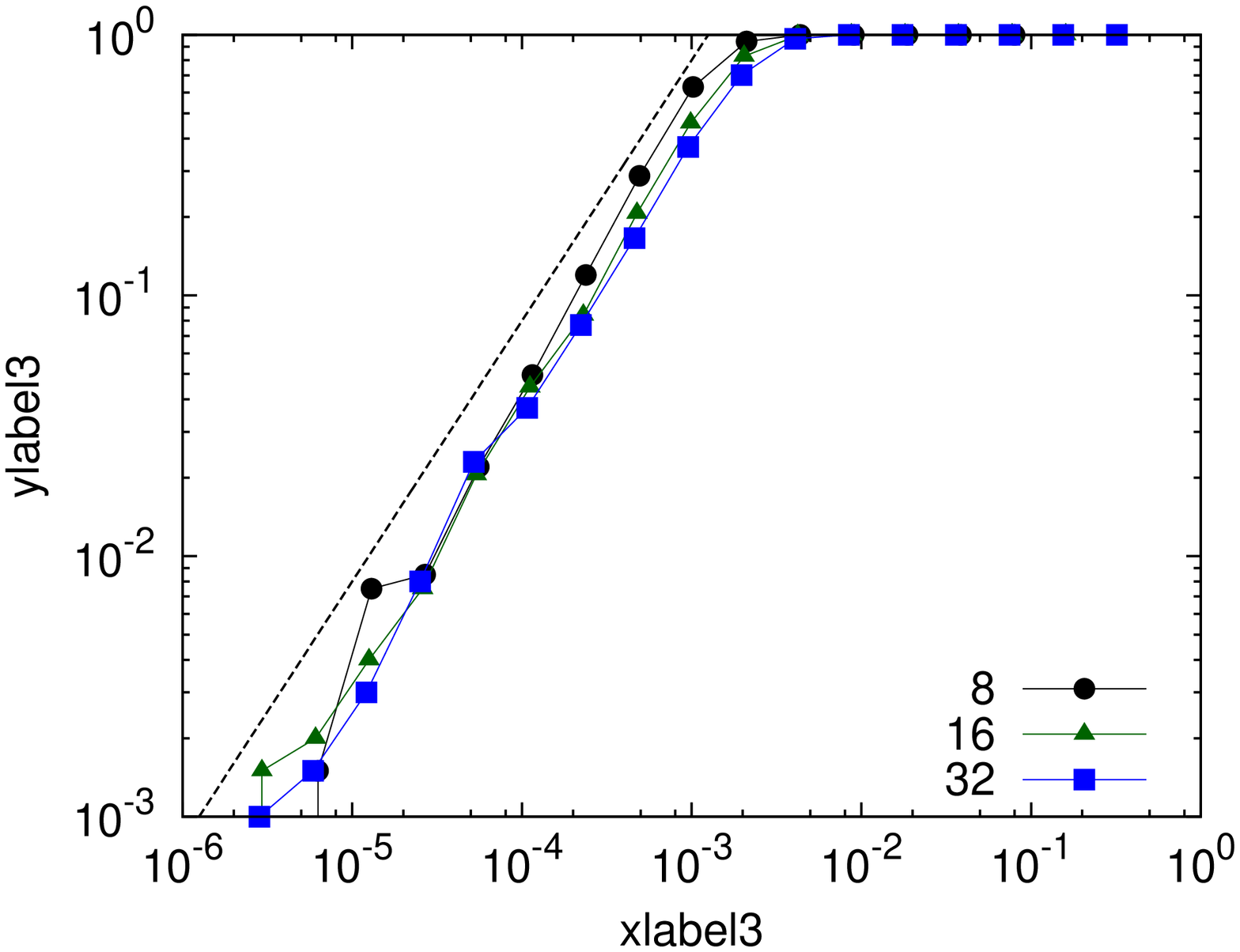}
\caption{The same as Fig.~\ref{fig:mc_nl} but for model C1.}
\label{fig:mc_nl_c}
\end{figure}

According to calculations of LEs we can distinguish 
chaotic and regular regimes, and calculate the probability 
of their appearance in models A, B, C.
The results for coupled symplectic maps of model A are presented 
in Fig.~\ref{fig:mc_kk}.
A typical lower cutoff for the LE calculated over time 
interval $T=10^8$ was $\approx 2.5\cdot 10^{-8}$, so we attributed 
all the realizations with $\lambda>5\cdot 10^{-7}$ to chaos. 
Defined in this way the total measure of initial conditions in the phase space that yield chaos $P_{ch}$
decreases with $K$ and $L$. The rescaled plot shows that for 
small $K$ and large $L$ the scaling relation 
\begin{equation}
 P_{ch}\sim K\cdot L
\label{eq:sckl}
\end{equation}
holds. The same  scaling $ P_{ch}\sim  K\cdot \beta$ 
is valid also for model B,
as demonstrated in Fig.~\ref{fig:mc_nl},
and for model C (Fig.~\ref{fig:mc_nl_c}).

The scaling with the system length $P_{ch}\sim L$ has been already 
discussed for model A in~\cite{Falcioni-Paladin-Vulpiani-89,Falcioni-91} 
and for disordered nonlinear lattices in~\cite{Pikovsky-Fishman-10}. 
It is based on the locality of chaos: the latter appears due to 
a local in space nonlinear interaction of localized modes, and 
not due to propagation of waves. Thus, in order to observe regularity 
in the whole lattice, the dynamics has to be regular in all subparts. 
Therefore, if the measure of chaos in a sublattice of length $L_0$,  
$P_{ch}(L_0)$, is small, then
$P_{ch}(L)\approx 1-(1-P_{ch}(L_0))^{L/L_0}$ from which 
the scaling $ \log P_{ch}\sim L$ follows.  An additional check of this relation is in Fig.~\ref{fig1dima}b below.

\subsection{Scaling of the Lyapunov exponent}

\begin{figure}[!hbt]
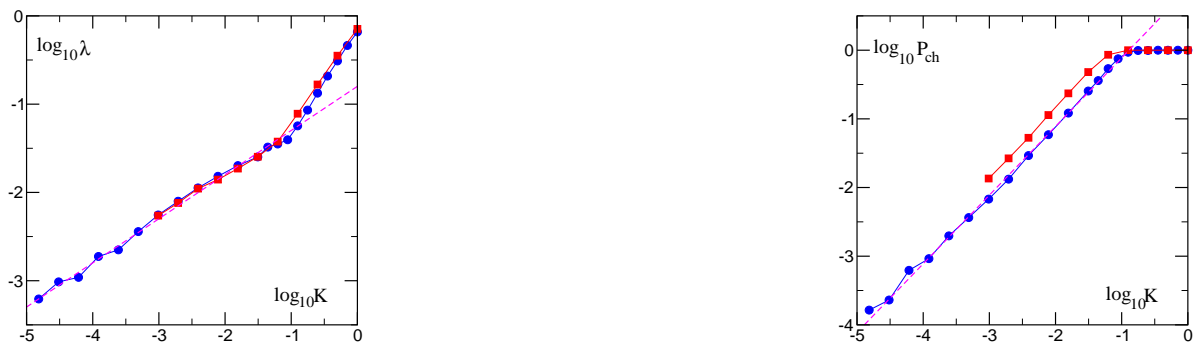

\centering
\includegraphics[width=0.30\textwidth]{fig1adima.eps}\hfill
\includegraphics[width=0.30\textwidth]{fig1bdima.eps}
\caption{Left panel: Dependence of the Lyapunov exponent
$\lambda$ on the perturbation parameter $K$ for model A at
$L=4$ (blue circles), $8$ (red squares).
Dashed straight line shows approximate dependence
$\lambda= 0.16 \sqrt{K}$ (for $N=4,8$). 
The fit of data gives the exponent of
the dependence $\lambda \propto K^a$
with $a=0.476 \pm 0.013$ (for $L=4$),
$0.450 \pm 0.012$ (for $L=8$) in agreement with the scaling 
(\ref{eq:scle}). 
Right panel: Dependence of the measure of chaos
$P_{ch}$ on the perturbation parameter $K$ for
$L=4$ (blue circles), $8$ (red squares).
Dashed straight line shows approximate dependence
$P_{ch}=7.75 K$ (for $L=4$);
for $L=8$ we find that $P_{ch}=14.32 K$
in agreement with the scaling (\ref{eq:sckl}). 
The fit of data gives the exponent of
the dependence $P_{ch} \propto K^b$
with $b=1.022 \pm 0.01$ (for $L=4$),
$1.036 \pm 0.01$ (for $L=8$). 
Up to $5 \times 10^5$
trajectories and time $t \leq 10^6$ have been used
to compute the Lyapunov exponent $\lambda$ and 
determine the number of chaotic trajectories with $\lambda >0$. 
Certain checks have been made with $t=5 \times 10^9$ and 100 trajectories.
}
\label{fig1dima}
\end{figure}

Next, we studied the scaling properties of the value of LE.
 As one can already see from Fig.~\ref{fig:lere_kk}, 
the positive LEs concentrate around a maximal value that decreases with $K$ 
and $\beta$. We have found (see Fig.~\ref{fig1dima}a) that this maximal value
 is roughly independent on the length of the system 
$L$ and scales with nonlinearity parameters $K$ and $\beta$ as 
\begin{equation}
\lambda\sim K^{1/2} \; .
\label{eq:scle}
\end{equation}

\begin{figure}[!hbt]
\centering
\psfrag{xlabel0}[c][c]{$\log_{10}(\lambda\cdot K^{-1/2})$}
\psfrag{ylabel0}[c][c]{$w/ (K\cdot L)$}
\psfrag{xlabel1}[c][c]{$\log_{10}(\lambda\cdot \beta^{-1/2})$}
\psfrag{ylabel1}[c][c]{$w/ (\beta\cdot L)$}
\psfrag{K}[c][c]{$\beta$}
(a)\includegraphics[width=0.45\textwidth]{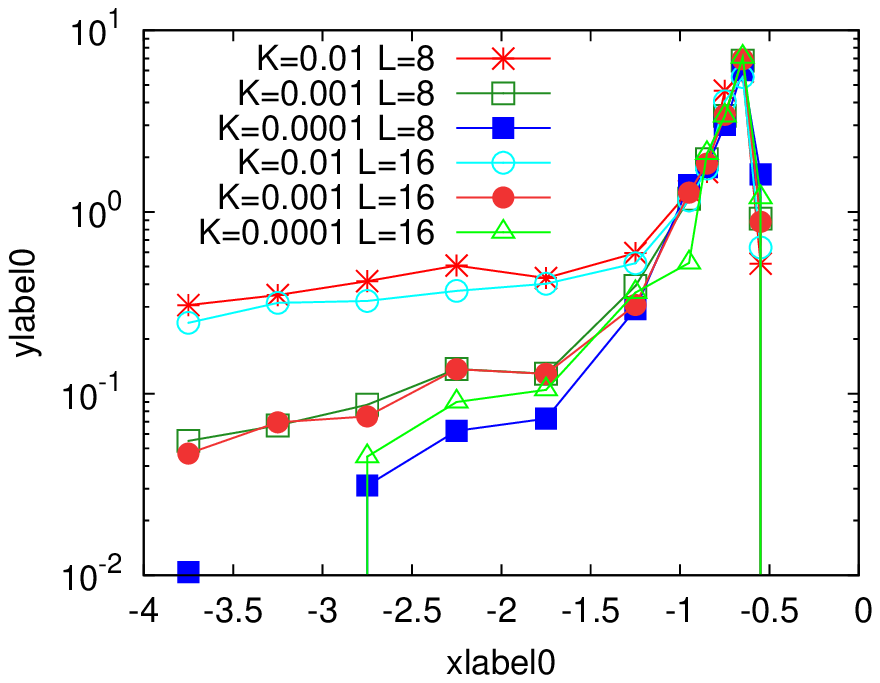}\hfill
(b)\includegraphics[width=0.45\textwidth]{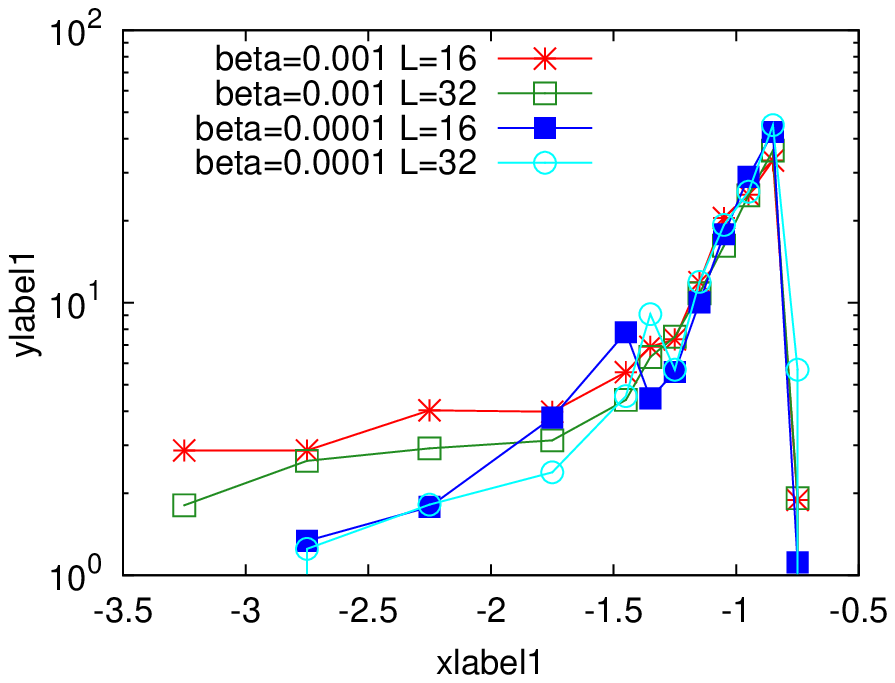}
\caption{(a): Distribution density of LEs in model A. (b): 
Distribution  density of LEs in model B.}
\label{fig:le-dens-0}
\end{figure}

\begin{figure}[!hbt]
\centering
\psfrag{xlabel0}[c][c]{$\log_{10}(\lambda\cdot K^{-1/2})$}
\psfrag{ylabel0}[c][c]{$W/ (K\cdot L)$}
\psfrag{xlabel1}[c][c]{$\log_{10}(\lambda\cdot \beta^{-1/2})$}
\psfrag{ylabel1}[c][c]{$W/ (\beta\cdot L)$}
\psfrag{beta}[c][c]{$\beta$}
(a)\includegraphics[width=0.45\textwidth]{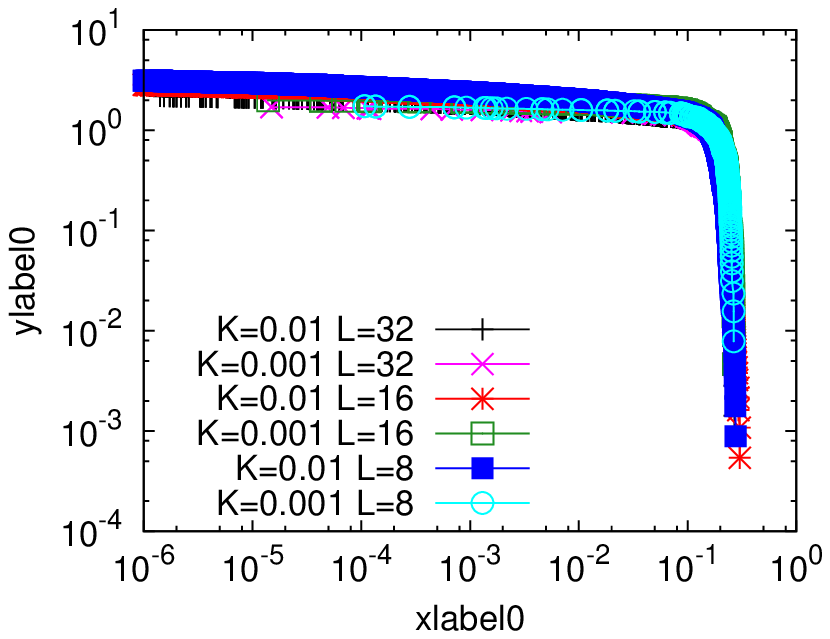}\hfill
(b)\includegraphics[width=0.45\textwidth]{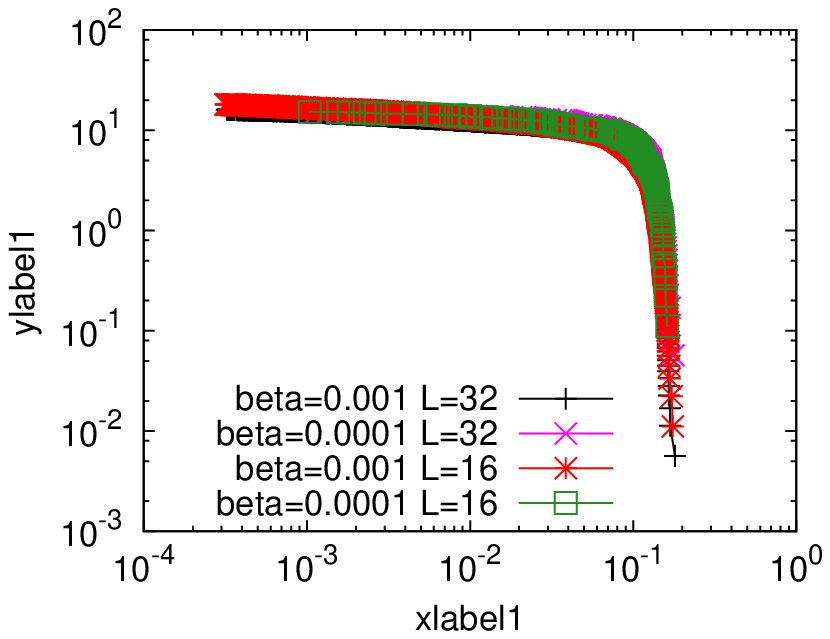}
\caption{(a): cumulative distribution of LEs in model A. (b): 
cumulative distribution of LEs in model B.}
\label{fig:le-dens}
\end{figure}

To demonstrate the scaling of the Lyapunov exponents we calculated 
their probability distribution densities $w(\lambda)$. Because of 
the relation $P_{ch}=\int_{\lambda_{th}}^\infty w(\lambda) \;d\lambda$ 
(where $\lambda_{th}$ is the cutoff value) the appropriate scaling 
for this density is that of $P_{ch}$, i.e. $K\cdot L$. According 
to (\ref{eq:scle}), the appropriate scaling of the argument of 
the density is $\lambda K^{-1/2}$.  We plot rescaled in this 
way distribution densities of LEs  
for models A and B in 
Figs.~\ref{fig:le-dens-0} and \ref{fig:le-dens}. We present here results 
for the distribution density $w$, for constructing of which 
some arbitrary bins have been used, and for a cumulative distribution
$W(\lambda)=\int_\lambda^\infty   w(\lambda) \;d\lambda$
where all data are presented, respectively. 
We note that the scaling law (\ref{eq:scle}) differs from the scaling 
$\lambda\sim K^{2/3}$ 
suggested in~\cite{Falcioni-Paladin-Vulpiani-89,Falcioni-91}. 
For the model B we find the same scaling relation 
$ \lambda\sim \beta^{1/2}$ as it is shown in Fig.~\ref{fig:le-dens}b.
For the model C we find the similar relation.

\subsection{Strong and weak chaos}
\begin{figure}[!hbt]
\centering
\psfrag{xlabel1}[c][c]{$K$}
\psfrag{ylabel1}[c][c]{$P_{wch}$}
\includegraphics[width=0.45\textwidth]{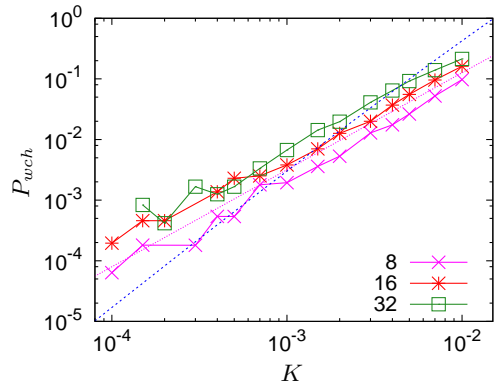}
\caption{Probability of ``weak chaos'' $P_{wch}$ with the LE between 
the low $\lambda>10^{-7}$ and high ($\lambda <0.03\cdot K^{1/2} $) 
cutoffs in Fig.~\ref{fig:le-dens} in model A. The pink dotted line is 
$P_{wch} \sim K^{1.6}$, the dashed curve corresponds to 
the estimate $P_{wch}\sim K^{2.5}(\ln K)^2$,
discussed at the end of Section \ref{sec:md}
in relation to the regime of fast Arnold diffusion 
analyzed
in \cite{Chirikov-Vecheslavov-97}.}
\label{fig:le-dens-1}
\end{figure}

\begin{figure}[!hbt]
\centering
\psfrag{xlabel0}[c][c]{time interval}
\psfrag{ylabel0}[c][c]{local LE}
\includegraphics[width=0.45\textwidth]{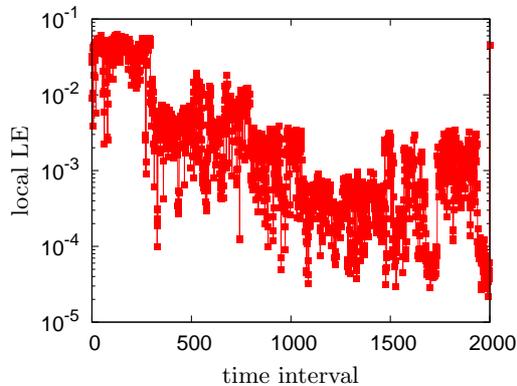}
\caption{Local LEs in model A as a function of time. 
Each value of LE is calculated
 over time interval of length  $10^6$, 
here $2\cdot 10^3$ such time intervals are shown; 
$K=0.05$, $L=8$.}
\label{fig:localle}
\end{figure}

There is also a substantial part of trajectories that 
have LEs between the lowest cutoff (determined by the averaging time) 
and the largest value $\sim \sqrt{K}$. We will distinguish 
these regimes by referring to the dynamics with LEs in 
the peak of distribution in 
Fig.~\ref{fig:le-dens-0} as \textit{strong chaos} 
while the dynamics with lower LEs will be called \textit{weak chaos}.
As it will be discussed later, it might be that
the regime of weak chaos is that where
 the fast Arnold diffusion
discussed in \cite{Chirikov-Vecheslavov-97} occurs.
We show in Fig.~\ref{fig:le-dens-1}
that the total probability $P_{wch}$ to observe this weak chaos scales as  
\begin{equation}
P_{wch} \propto K^{\nu_{wch}} L \; , \;\;\; \nu_{wch} \approx 1.6 \; .
\label{eq:wch}
\end{equation}
In Fig.~\ref{fig:localle} we show an example of a local 
in time LEs for one long trajectory in model A.
It shows existence of transitions between regimes with 
strong chaos and weak chaos.

\section{Resonances as a source of chaos}
\label{sec:res}

\begin{figure}[!hbt]
\centering
\psfrag{xlabel0}{$d/\sqrt{K}$}
\psfrag{ylabel0}{$P_{ch}$}
\psfrag{xlabel1}{$d/\sqrt{\beta}$}
\psfrag{ylabel1}{$P_{ch}$}
(a)\includegraphics[width=0.35\textwidth]{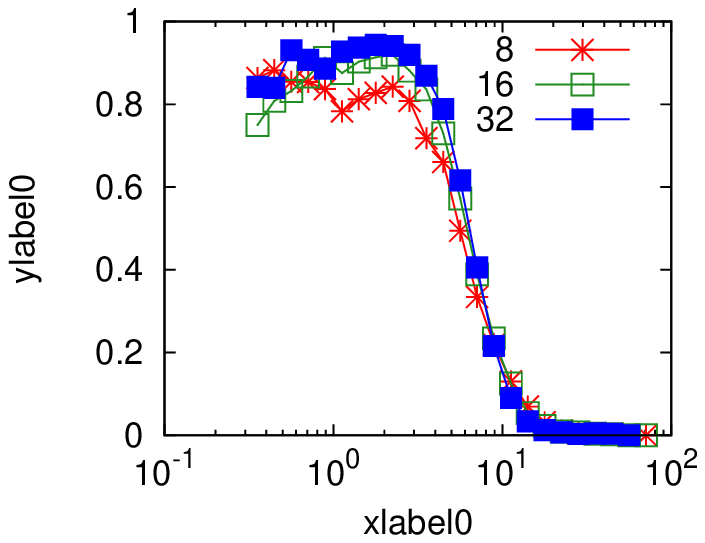}\hspace{4em}
(b)\includegraphics[width=0.35\textwidth]{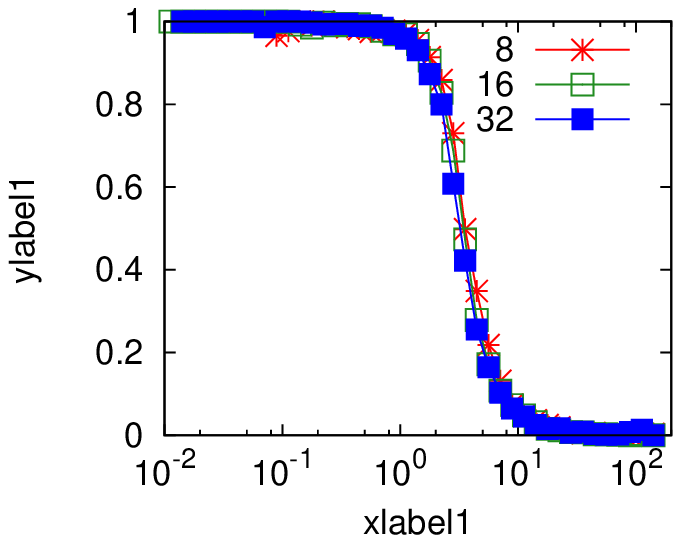}
\caption{The probability to observe  chaos in dependence on 
the resonance in initial data for models A (a) and B (b).  
The data for different $L$ and $K,\beta$ collapse if 
the distance in initial frequencies is scaled according 
to $\sqrt{K}$ or $\sqrt{\beta}$.}
\label{fig:res_kk}
\end{figure}

In order to characterize conditions under which chaos occurs at very small
coupling, we have looked on resonances, and have found that chaos 
is highly correlated with the triple resonance at which 
the frequencies of three neighboring oscillators nearly coincide. 
For models A and B we illustrate this in Figs.~\ref{fig:res_kk}, respectively. 
Here the probability of chaos $P_{ch}$ is shown vs. renormalized distances 
of initial frequencies of oscillators. For model A we have defined 
this distance as
$d=\text{min}_k[(f(p^{(0)}_k-p^{(0)}_{k+1})^2+
(f(p^{(0)}_{k+1}-p^{(0)}_{k+2}))^2]$. 
Here $f(x)=2|\sin 0.5x|$ measures the closeness of 
two initial momenta modulo $2\pi$. A small value of $d$ indicates 
that somewhere in the lattice three initial nearby momenta $p^{(0)}$ 
are close to each other. Then, for different realizations of 
initial conditions, different $K$ in the range $[0.001,0.2]$ and 
different lattice lengths $L=8,16,32$ we determined the probability 
for chaos to occur vs. $d/\sqrt{K}$. One can see that for 
different lattice lengths the curves are close to each other, 
thus indicating that indeed the occurrence of resonances 
is a necessary prerequisite for chaos. In a similar analysis 
for model B we used
$d^2=\text{min}_k[(\sqrt{p_k(0)}-
\sqrt{p_{k+1}(0)})^2+(\sqrt{p_k(0)}-\sqrt{p_{k-1}(0)})^2]$.

In Fig.~\ref{fig:res_kk} we demonstrate the correlation between 
the occurrence of resonance (small $d$) and the probability 
to observe chaos $P_{ch}$. Moreover, we see here the scaling that 
in fact $d$ should be compared with $\sqrt{K}$ (or $\sqrt{\beta}$ for model B). 

The physical reason for the scaling results presented in 
previous sections is the following (for simplicity of presentation, 
we refer here to model A only, the same arguments work for models B and C).
There is a finite probability that
three nearby particles will have their frequencies
$\omega_i=p_i$ close to each other, within
the frequency range $\Delta\omega=\sqrt{K}$.
The probability of such an event is
$P \sim K$, since the first particle may have any frequency,
the probability to have the second in the range $\sqrt{K}$
is $\sqrt{K}$ and the probability to have the third
in the same range is also  $\sqrt{K}$. This gives the
probability of the resonance $P \sim K $ for a lattice with three
particles and $P \sim K L$ for a chain with L oscillators. 
Similar arguments work for models B,C.
It is important to note that
in the case of such a 3-particle resonance,
the KAM arguments are not valid and the dynamics remains
chaotic at arbitrary small perturbation $K$.
The situation is similar to the one considered in
\cite{Chirikov-Shepelyansky-82} where three linear oscillators with the same
frequency remain chaotic at arbitrary small
nonlinear coupling between them. Indeed, in our case
the numerical analysis shows that almost all chaotic
trajectories (those with positive Lyapunov exponent)
have three nearby particles with close frequencies.

\begin{figure}[!hbt]
\centering
\psfrag{xlabel0}{$\phi_2$}
\psfrag{xlabel1}{$\phi_2$}
\psfrag{ylabel0}{$J_2$}
\psfrag{ylabel1}{$J_2$}
(a)\includegraphics[width=0.35\textwidth]{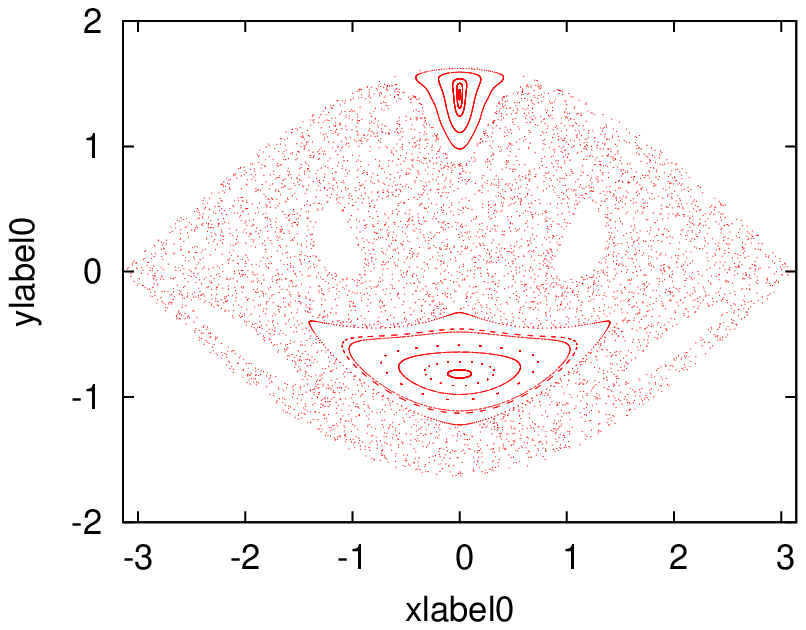}\hspace{5em}
(b)\includegraphics[width=0.35\textwidth]{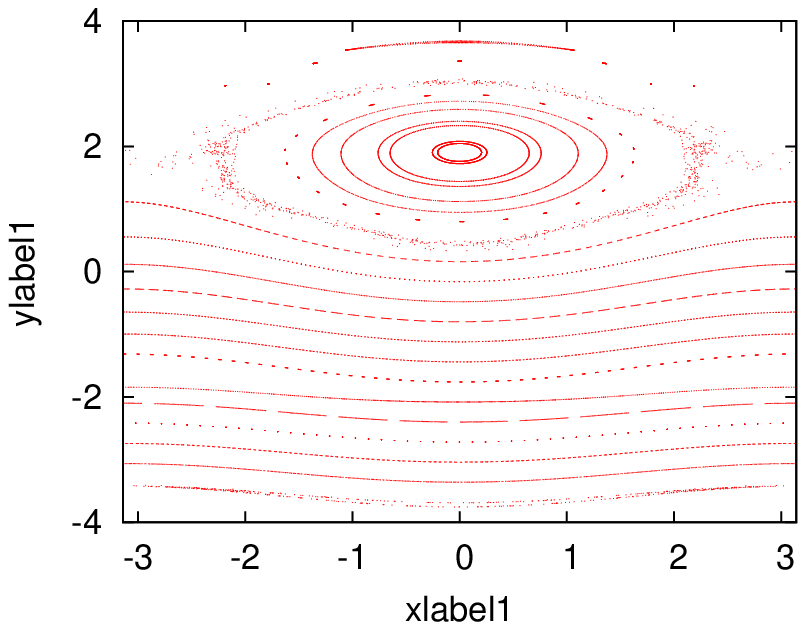}
\caption{Poincar\'e sections of variables $\phi_2,J_2$ 
for the resonance Hamiltonian (\ref{eq3}) at $\phi_1=0$. (a): $H_R=0$,
here chaos is dominant.
(b): $H_R=10$, here the dynamics is typically quasiperiodic.}
\label{fig:3ham}
\end{figure}

To understand this phenomenon in a better way
let us consider the case when initially
at three neighboring sites the values of actions $p_i$
are close to their average value $P=(p_1+p_2+p_3)/3$.
Then the evolution of these three particles, considered separately 
from the rest (what can be justified by arguing that 
nonresonant terms effectively disappear after averaging) 
is described by the mapping
\begin{align}
\bar{p}_1-p_1=K\sin(x_2-x_1)\;,&\quad \bar{x}_1-x_1=\bar{p}_1\;,\\
\bar{p}_2-p_2=K\sin(x_1-x_2)+K\sin(x_3-x_2)\;,&\quad \bar{x}_2-x_2=\bar{p}_2\;,\\
\bar{p}_3-p_3=K\sin(x_2-x_3)\;,&\quad \bar{x}_3-x_3=\bar{p}_3\;.
\end{align}
Exploring the integral $p_1+p_2+p_3=const$ and 
performing a canonical transformation to new conjugate 
coordinates according to
\[
\phi_1=x_1-x_2,\;\phi_2=x_3-x_2,\;\phi_3=x_1+x_2+x_3,\;
p_1=I_1+I_2+P,\;p_2=-I_1-I_2+I_3+P,\;p_3=I_2+I_3+P,
\]
we obtain a two-dimensional mapping
\begin{align}
\bar{I}_1-I_1=-K\sin\phi_1\;,&\quad \bar{\phi}_1-\phi_1=2\bar{I}_1+\bar{I}_2\;,\\
\bar{I}_2-I_2=-K\sin\phi_2\;,&\quad \bar{\phi}_2-\phi_2=2\bar{I}_2+\bar{I}_1\;,
\end{align}
which due to smallness of $I_1,I_2$ and of $K$ can be approximated 
as a continuous-time system with Hamiltonian 
$H=I_1^2+I_2^2+I_1I_2-K\cos\phi_1-K\cos\phi_2$.
After rescaling of actions
to $J_i=I_i/\sqrt{K}$ and time to $\tau=\sqrt{K} t$
we come to dimensionless resonance Hamiltonian
\begin{equation}
\label{eq3}
H_R(J_1,J_2,\phi_1,\phi_2)= J_1^2+J_2^2+J_1J_2 - \cos(\phi_1) - \cos(\phi_2)
\end{equation}
Note that this rescaling proofs the dependencies $\sim K^{1/2}$ for the 
allowed deviations from the resonance condition. Also the rescaling 
of time proofs the scaling of the Lyapunov exponent with $K$ according 
to (\ref{eq:scle}). 

According to the Chirikov resonance-overlap criterion \cite{Chirikov-79}
the dimensionless dynamics of Hamiltonian (\ref{eq3})
is chaotic for small values of energy (i.e. close to resonance) 
and chaos disappears if the energy is large. The Poincar\'e sections
for $H_R$ for $H_R=0$ and $H_R=10$ are shown in Fig.~\ref{fig:3ham}
confirming this picture.

\section{Properties of diffusion and weak chaos}
\label{sec:md}

While LEs serve as an important indication for chaos, other quantities 
like correlations are important to characterize irregularity of the dynamics. 
For the Chirikov standard map an important statistical quantity is 
the diffusion constant of the momentum $p$: at large times $T$ the dynamics of 
$p$ can be considered as a random walk with a diffusion constant $D$ 
defined according to $\langle (p(T)-p(0))^2\rangle = D T $. 
For the Chrikov standard map the dependence of $D$ on the parameter $K$ 
is known in detail~\cite{Chirikov-79,Lichtenberg-Lieberman-92}. 

\begin{figure}[!hbt]
\centering
\psfrag{xlabel0}[c][c]{$T$}
\psfrag{ylabel0}[c][c]{Variance}
\psfrag{xlabel1}[c][c]{$K$}
\psfrag{ylabel1}[c][c]{$D$}
\includegraphics[width=0.45\textwidth]{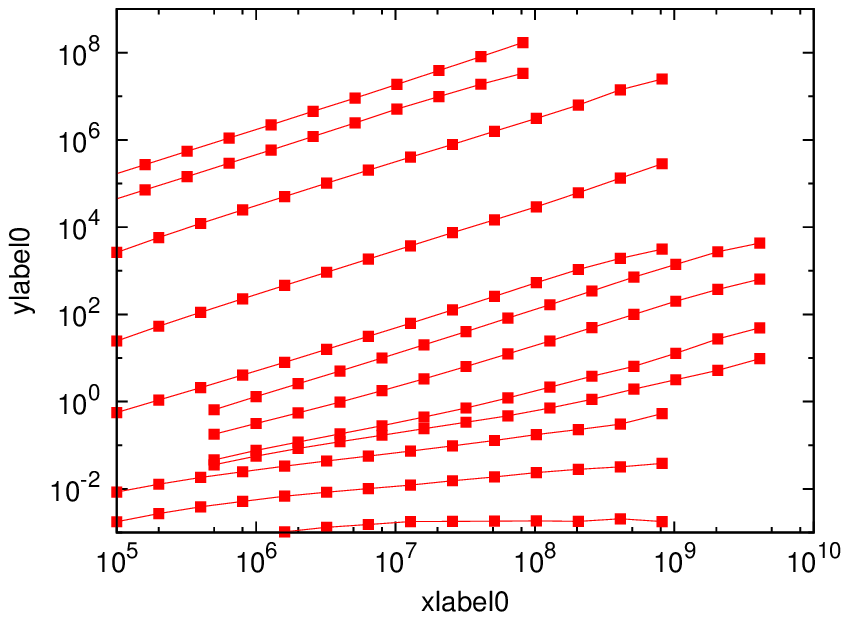}\hfill
\includegraphics[width=0.45\textwidth]{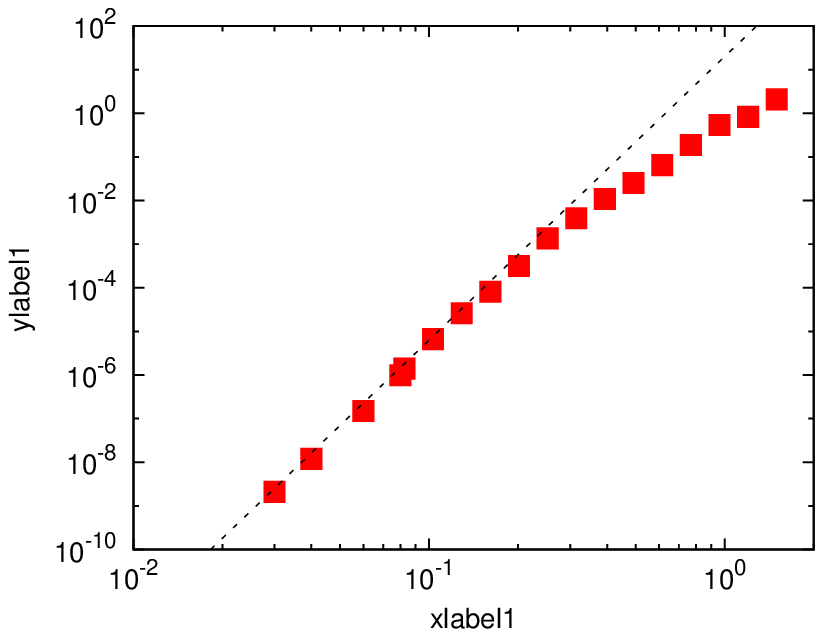}
\caption{Left panel: Variance $\langle (p(T)-p(0))^2\rangle$ 
as a function of time $T$ calculated in a lattice of length $L=64$. 
From top to bottom:
$K=1.5,\;1.,\;0.5,\;0.2,\;0.11,\;0.08,\;0.06,\;
0.05,\;0.04,\;0.03,\;0.01,\;0.005,\;0.002$. Right panel: dependence of 
the diffusion constant on $K$ in range $0.03<K<1.5$. Dashed line shows 
relation $D=20K^{6.5}$.}
\label{fig:diff}
\end{figure}

For the coupled symplectic maps (model A) numerical 
computations~\cite{Kaneko-Konishi-89,Konishi-Kaneko-90}, 
performed in a range $0.1<K<1$, indicated 
a weak diffusion  at $K=0.1$, the authors fitted the data with 
a stretched exponential dependence. 
Here we extend these calculations and show the results in Fig.~\ref{fig:diff}. 
One can see a strong decrease of the diffusion constant with $K$, 
which for small $K$ is close to a power-law dependence 
\begin{equation}
D\sim K^{\nu_D} \; , \;\; \nu_D \approx 6.5 \; .
\label{eq:difdk}
\end{equation}
A similar value of the exponent was obtained from the statistics of Poincar\'e
recurrences in the range $0.1 \leq K \leq 1$ \cite{Shepelyansky-10}.  
We note that for  model C the above equation implies
$D \propto \gamma^{\nu_D}$.
The value of the exponent $\nu_D$ is close to the value
given by Chirikov and Vecheslavov 
\cite{Chirikov-Vecheslavov-93,Chirikov-Vecheslavov-97}.
However, they calculated the diffusion indirectly by expressing it
via an effective width $w_s$ of a separatrix layer of a nonlinear resonance
with the additional relation $D \sim K^{3/2} w_s^2$, which was verified 
with the direct computations of the Arnold diffusion 
in systems with a few degrees of freedom. In fact, the value of
$w_s$ is determined in  \cite{Chirikov-Vecheslavov-93,Chirikov-Vecheslavov-97}
via the computation of the period of oscillations around 
a separatrix layer of a nonlinear resonance that is 
 related to the computation
of LE. Due to this indirect method,  Chirikov and Vecheslavov
were able to obtain the variation of the Arnold diffusion constant
$D_A$ by 50 orders of magnitude! On a scale of first 30 orders of magnitude
the decay of the diffusion constant
$D$ is well described by the power law with
 $\nu_D =6.5$ (see Fig.~1 in \cite{Chirikov-Vecheslavov-97}).
The main message of these amazing calculations is a non-exponential
decay of  $D$, and hence of the chaos measure $w_s$,
with the decrease of nonlinearity parameter $K$.
This result is in a drastic difference from the asymptotic
Nekhoroshev-like estimates based on the KAM theory 
\cite{Nekhoroshev-77,Lochak-92} which give
exponential decrease of $D$ and $w_s$
as $K \rightarrow 0$. Of course, there is no
formal contradiction since the results for
fast Arnold diffusion \cite{Chirikov-Vecheslavov-97}
are always obtained at small but finite 
$K$ values. However, an algebraic decrease with $K$
indicates on an existence of weak chaos component
with relatively large measure. The heuristic arguments
for this phenomenon were presented in \cite{Chirikov-Vecheslavov-97}.
According to the results of \cite{Chirikov-Vecheslavov-97} one has
for model A:
\begin{equation}
D\sim K^{3/2} w^2_s\; , \;\; w_s \sim K^{\nu_s} \; , \; \nu_s \approx 2.5 \; ,
\; \nu_D =2 \nu_s+3/2 \; ,
\label{eq:difchiv}
\end{equation}
for $K > 1.6 \cdot 10^{-5}$. Here, $w_s$ is a dimensionless measure
of the chaotic separatrix layer of the resonance between 
two nearby oscillators. For the range 
$2 \cdot 10^{-6} < K \leq 1.6 \cdot 10^{-5}$ the decay of $D$
is compatible with the power law $D \propto K^{15}$
but this range of $K$ variation is not very large.
The global dependence $D(K)$ is fitted by the dependence of Eq.~(5.8) in
\cite{Chirikov-Vecheslavov-97} which however has no complete theoretical
explanation.

The reason why one can hardly compute the diffusion coefficient at smaller $K$ 
is clear from the inspection of the dependence of the variance on time in 
Fig.~\ref{fig:diff}. For small $K$ one observes a normal diffusion 
only when the variance exceeds $\approx 1$, below this value 
the diffusion looks like anomalous one with the variance proportional 
to a power of time. This means that a ``random walk'' inside 
the periodicity cell $[0,2\pi)$ is highly correlated, 
while only cell-to-cell walk demonstrates a normal diffusion. 
For small $K$ the mean first passage time to the next cell 
becomes extremely large -- nearly $10^9$ for $K=0.03$, while for $K<0.03$ this  
mean passage time is of order or larger than the total 
integration time and only the anomalous diffusion is observed.

The obtained properties of diffusion should be contrasted to 
the properties of LEs, as both quantities give some characteristic times 
of the system. We have demonstrated that these times become extremely 
different for small non-integrabilities, as the Lyapunov exponent 
$\lambda \sim K^{1/2}$ decreases rather weakly with $K$
while the diffusion constant 
$D \sim K^{6.5}$  drops much more rapidly. We interpret this as indication 
that chaos is mainly ``local'', not leading to large deviations of variables. 
This picture corresponds well to the discussed above effective resonances 
as the origin of chaos: in the triple resonance described above in 
Section~\ref{sec:res}, the sum of all momenta is a conserved quantity, 
so that the chaotic dynamics like in Fig.~\ref{fig:3ham} does not lead 
to a large deviation of momenta involved in the resonance.
Indeed, there is strong chaotic dynamics inside the triplet resonance,
but the sum of three resonant actions is a constant 
in the resonance approximation
that would give a zero diffusion coefficient $D=0$.
However, the resonant approximation is not exact and 
it is destroyed by nonresonant terms and higher order
perturbations that leads to a finite value of the diffusion
$D \sim K^{6.5}$. A mixture of strong chaos,  which is however bounded
due to an additional integral of motion, and
a slow but unbounded diffusion produced by  weak chaos
makes the numerical computation of the diffusion rate
a rather difficult task.
In fact,  usual very  powerful methods
discussed in \cite{Chirikov_etal-85}, which allowed to compute
as small diffusion rate as  $10^{-22}$,
are not working in such a situation
and only computations at very long times
allow to determine directly the value of $D$.

The physical origins of the power law decay of the diffusion rate
with $K$ (\ref{eq:difdk}) are still to be understood.
The theoretical heuristic arguments presented in 
\cite{Chirikov-Vecheslavov-97} assume that in the regime of weak chaos
a trajectory follows mainly those chaotic resonant layers which
have locally most large size. An optimization
over various resonances leads to a certain power low decay
for $w_s$ and $D$ which gives $\nu_s = e=2.718...$
and $\nu_D=1.5+2e = 6.936...$ respectively
(we remind that for the Chirikov standard map 
$w_s \propto \exp(-\pi^2/\sqrt{2K})$ 
\cite{Chirikov-79,Chirikov-Vecheslavov-97}).
This theoretical value of the exponent $\nu_D$
is in a satisfactory agreement with the numerical value 
found at not very small $K$ values.
However, at very small values of $K < 10^{-5}$ such arguments 
should be modified to fit an unknown dependence
of resonance amplitudes in high orders of perturbation theory
\cite{Chirikov-Vecheslavov-97}. According to the heuristic arguments
\cite{Chirikov-Vecheslavov-97}
the main contribution to diffusion is given by the resonances with
an effective resonance harmonic numbers $\tilde{M}_0 = \ln (1/\sqrt{K})$
with a dimensionless measure of chaos inside one given
resonance separatrix layer $w_S$.  We may argue that
the number of such layers grows with 
 $\tilde{M}_0$ at least as $\tilde{M}_0^2$  
so that the total measure of weak chaos
can be estimated as 
$P_{wch} \propto  \tilde{M}_0^2 w_s \propto (\ln K)^2 K^{2.5}$.
This dependence is in a satisfactory agreement
with the data of Fig.~\ref{fig:le-dens-1} 
(see the dashed curve there)
and the empirical exponent value
$\nu_{wch} \approx 1.6$ in (\ref{eq:wch}).
Thus we can say that our data for the measure of weak chaos
are in a satisfactory agreement with the 
numerical results~\cite{Chirikov-Vecheslavov-97}.

On the other hand, the origin of such a weak chaos component 
is still to be clarified. Indeed, the studies and arguments presented in 
\cite{Chirikov-Vecheslavov-97} did not take into account the 
strong chaos based on triplet resonances which
exists at arbitrary small $K$. This strong chaos component
emerges as the result of triple primary resonances but it is clear
that a similar mechanism can work for higher order resonances
which may be at the origin of the weak chaos component.
On the other hand, the triple-like resonances of higher order
in $K$ should lead to appearance of a certain number of trajectories
with the LEs $\lambda \propto K^{m/2}$ with $m \geq 2$
that is, however, is not visible in the distribution
of LEs in Figs.~\ref{fig:le-dens-0},\ref{fig:le-dens},\ref{fig:le-dens-1}.
It is however, possible that other tiny chaotic layers
hide such contributions. 
Further studies are required to clarify these points
especially in the regime with large $L \gg 15$. An indication on 
the complex internal structure of weak chaos provides 
Fig.~\ref{fig:localle} above, which demonstrates 
how a trajectory visits regions with different LEs along a very long evolution.

\section{Spreading of chaos}
\label{sec:spread}

Above we discussed the local properties of chaos
computing the Lyapunov exponents and the diffusion rate
in the regime when all nonlinear oscillators are 
populated in the initial state. Another type of question appears
for the model C2 (\ref{eq:modc}) when only one of few nearby oscillators
are initially excited with the total energy $E_{tot}=1$
and $\gamma=1$ while all other oscillators have zero energy. 
Since the total energy is conserved
we face the question on a possibility of energy spreading
over the whole lattice of size $L$. This is related to the question
of ergodicity of large finite lattices
at small energies. In the case when both nonlinear terms in 
the Hamiltonian (the local potential and the coupling) 
have the same power (e.g. the coupling has
 power $4$ instead of $6$, such a model can be called model C44) 
then it is known that 
a thermalization takes place at arbitrary small total energy according to
the arguments given in \cite{Ahnert-10}. Of course, the time
for such global ergodicity grows as a power of system size $L$.
For models with a nonlinear destruction of the 
Anderson localization,
we have the terms with powers $2$ for local potential and $4$ for coupling
in (\ref{eq:modc}), which we call model C24. In this case 
it is found that a slow subdiffusive
spreading over the lattice takes place 
up to very long times $t \sim 10^9$ 
(see details in recent papers
\cite{Pikovsky-Shepelyansky-08,Garcia-Mata-Shepelyansky-09,%
Flach-Krimer-Skokos-09,%
Flach-10,Laptyeva-etal-10,Mulansky-Pikovsky-10,%
Johansson-Kopidakis-Aubry-10,Krimer-Flach-10,%
Ahnert-10}).
The model C2 corresponds to a new situation 
for energy spreading when
the unperturbed integrable Hamiltonian is nonlinear 
and the coupling between nonlinear modes 
has higher nonlinearity. In contrast to the FPU problem, here
the coupling between modes is local and the randomness
in local nonlinear frequencies $\eta_k$ excludes
any proximity to a full hidden  integrability.
\begin{figure}[!hbt]
\centering
\psfrag{xlabel1}[c][c]{time $t$}
\psfrag{ylabel1}[c][c]{$(\Delta k)^2$}
\psfrag{xlabel2}[c][c]{$t$}
\psfrag{ylabel2}[r][c]{$P$}
\includegraphics[height=0.45\textwidth,angle=270]{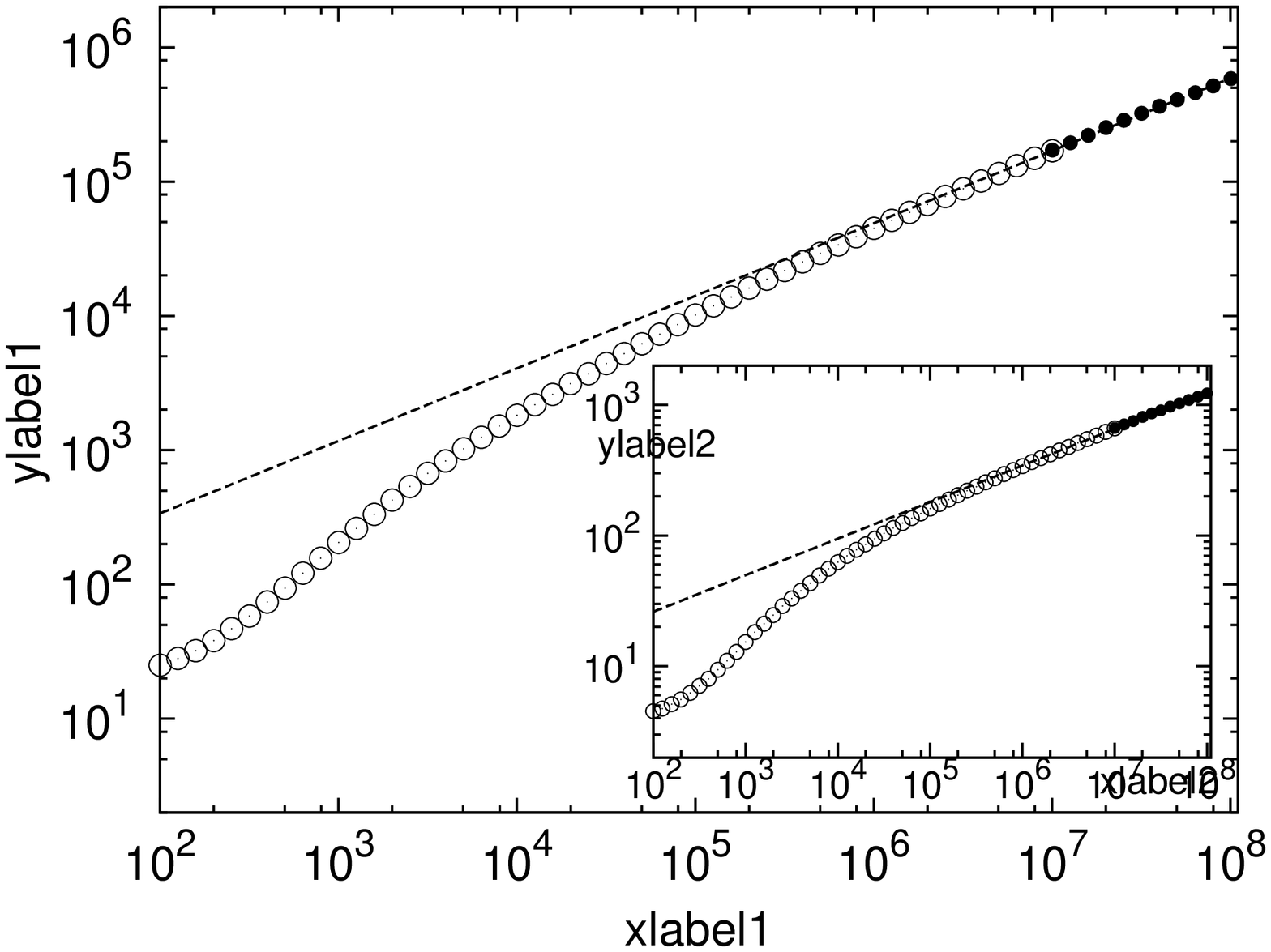} \hfill
\includegraphics[height=0.45\textwidth,angle=270]{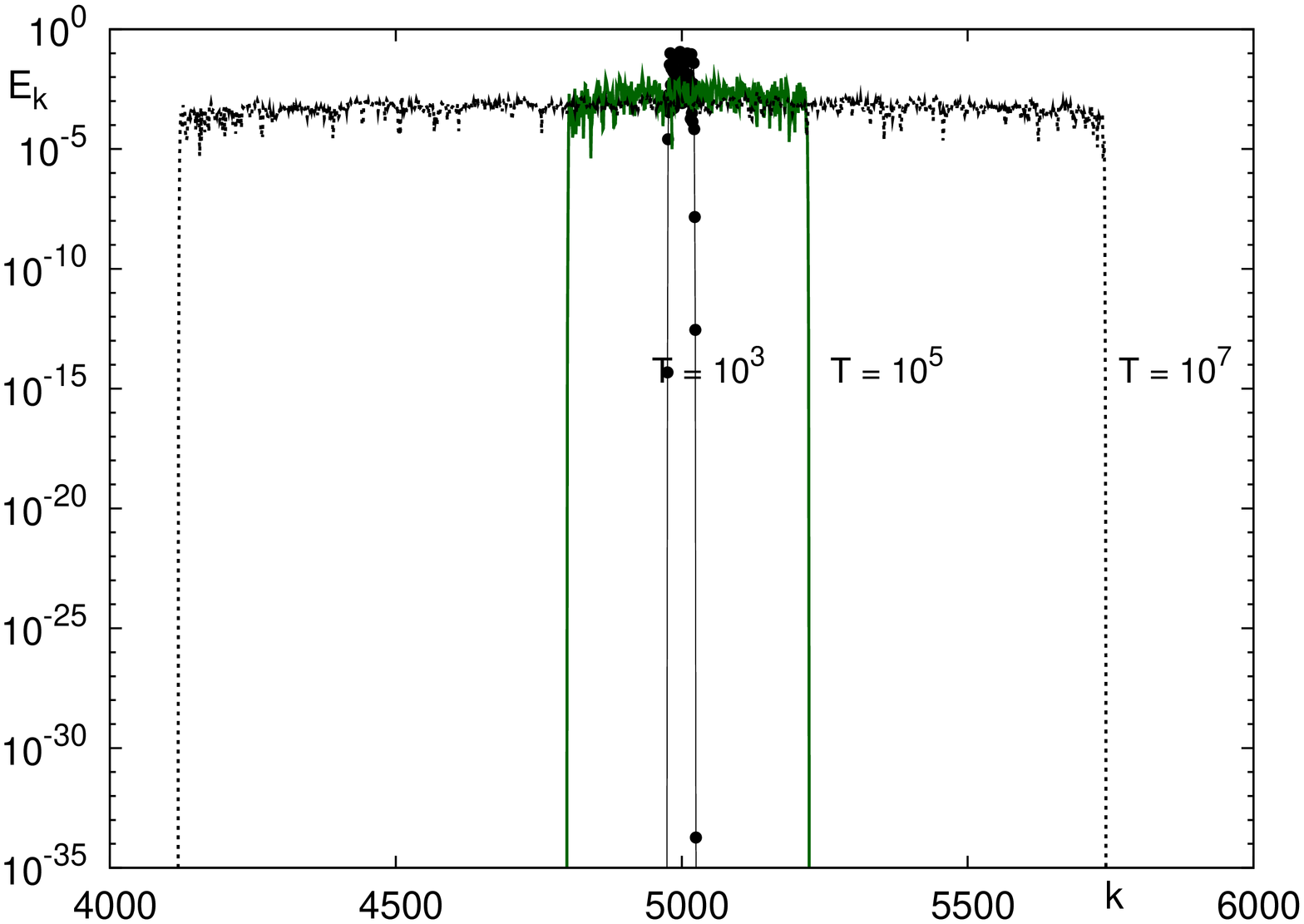}
\caption{Left panel: Spreading of the second moment 
$(\Delta k)^2= \sum_k (k-k_0) E_k/\sum_k E_k$ 
and participation number $P = (\sum_k E_k)^2/\sum_k E_k^2$ (inset) 
vs.\ time for initial single site excitation
$k_0=0$ in model C2.
Data up to $t=10^7$ (empty circles) has been averaged over 1000 
realizations of disorder and logarithmic time windows.
Long-time values until $t=10^8$ (full circles) were averaged over 24 realizations.
Dashed lines show subdiffusive growth $(\Delta k)^2  \sim t^\alpha$ 
and $P \sim t^{\alpha/2}$ where the fit of 
the asymptotic behavior ($t>10^5$) gave $\alpha=0.55 \pm 0.01$.
Right panel: energy distribution $E_k$ 
in model C2 at certain moments of time $t=10^3$, $10^5$, $10^7$ 
for one specific realization of disorder.}
\label{fig:spread}
\end{figure}

Let us assume that in model C2 with the above local
initial conditions
the energy spreads over the whole lattice of $L$ oscillators
with an approximate energy equipartition over $L$ sites.
After a rescaling of variables of this final state to a new time
$\tau \rightarrow L^{1/4} t$ we come to the model C2
with $\gamma \sim 1/\sqrt{L}$ and a
homogeneous initial condition, discussed in the previous sections.
In general, the probability of strong chaos in such a case scales as 
$P_{ch} \propto \gamma L \propto \sqrt{L}$ so that
we expect local strong chaos to occur almost surely 
in a sufficiently long
lattice. The same it true for the probability of weak chaos
even if in this case the sum value of the exponents in $L$
is close to zero. Although the probability to observe chaos
is high, it is important to note that this chaos is mainly local:
some modes are chaotic, e.g. triplets discussed above,
but other modes  generally oscillate nearly quasiperiodically. 
Indeed, in a system with many degrees of freedom
some modes can be chaotic while others can be close to integrable ones, 
without any contribution to the maximal LE.
Thus, it is not obvious if the local strong chaos can allow 
spreading from initial local state over the whole lattice.

Let us present here simple estimates on the possible rate of such 
a spreading using results for the diffusion in the weak chaos component. 
We assume that a chaotic spreading populates
the number of modes $N$ at time $t$.
Using rescaling given above we can argue that the
new mode $N+1$ will be populated due to the weak chaos diffusion
after a time scale 
$t_s/N^{1/4} \sim 1/D(\gamma) \sim \gamma^{-\nu_D} \sim N^{\nu_D/2}$.
This gives us an effective local diffusion rate in $N$ with
$N^2/t \sim 1/t_s \sim 1/N^{(2\nu_D+1)/4}$ leading to the subdiffusive
growth of the second moment $N^2$:
\begin{equation}
N^2 \sim t^{\alpha} \; , \;\; \alpha=8/(9+2\nu_D) \; .
\label{eq:spread}
\end{equation}  
For $\nu_D=6.5$ we obtain $\alpha=0.3636$. However, our results for spreading,
shown in Fig.~\ref{fig:spread}, give approximately $\alpha = 0.55$
that corresponds to $\nu_D \approx 2.77$.
We explain this difference in the following way.
At the maximum time $t_{max} =10^8$, reached in our numerical simulations,
the energy spreads over a number of modes 
$N \sim t_{max}^{\alpha/2}$ so that we have an effective
$\gamma \sim 1/\sqrt{N} \sim 0.02$ which is only at the beginning of the
decay with the exponent $\nu_D$ shown in Fig.~\ref{fig:diff}(right panel),
if we assume a simple relation
$\gamma=K$, which however still may have an additional numerical factor.
It is interesting to note that the case with $\nu_D=0$
corresponds to independence of $D$ on $N$ after rescaling
that is the case for nonlinear model C44 (with both potentials having power 4
in (\ref{eq:modc})) where the spreading goes indeed with the exponent
$\alpha=8/9$ as it is shown in \cite{Ahnert-10}.

\begin{figure}[!hbt]
\centering
\psfrag{xlabel1}[c][c]{$\omega$}
\psfrag{ylabel1}[c][c]{$D_{zk}$}
\includegraphics[height=0.3\textwidth,angle=270]{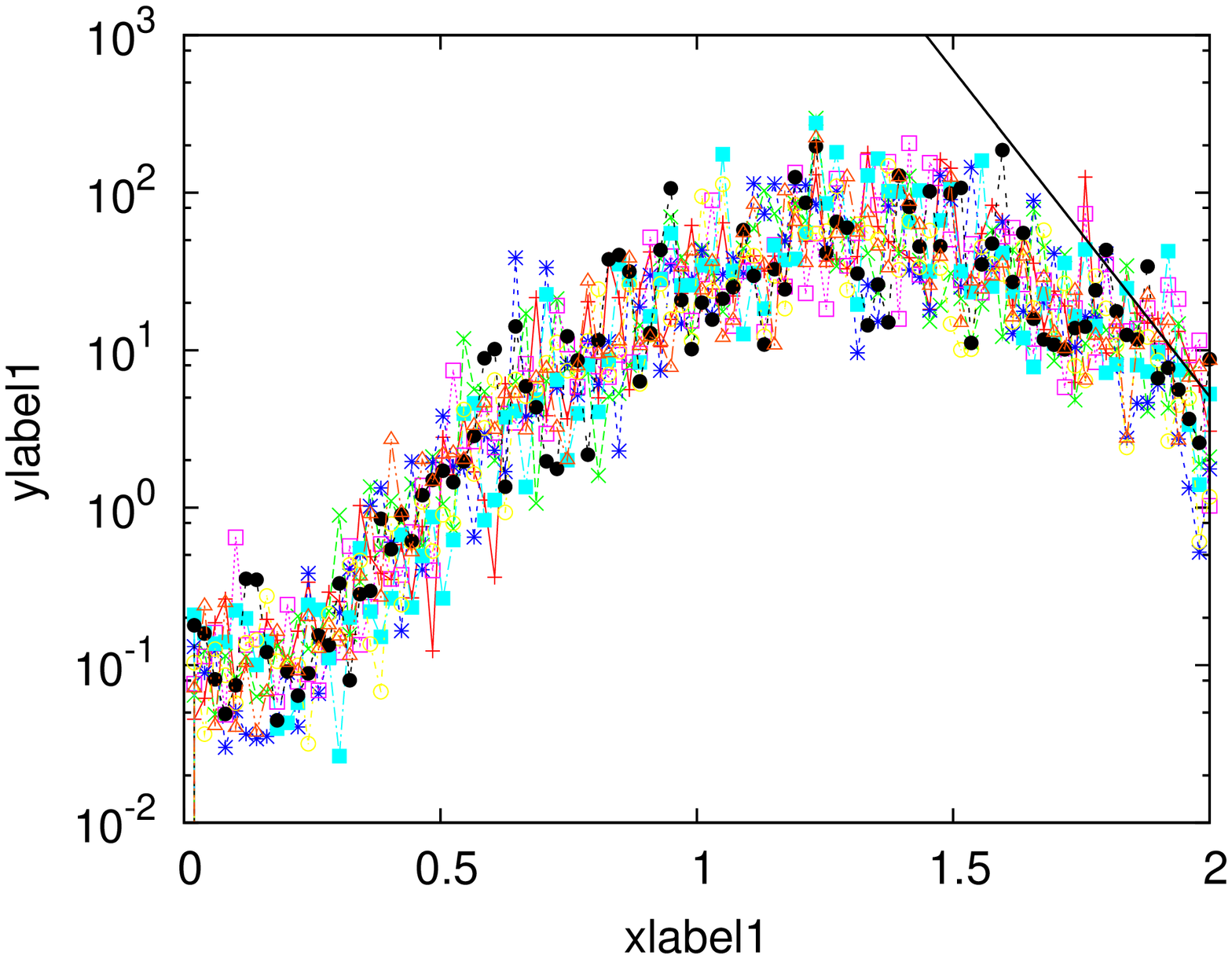}
\includegraphics[height=0.3\textwidth,angle=270]{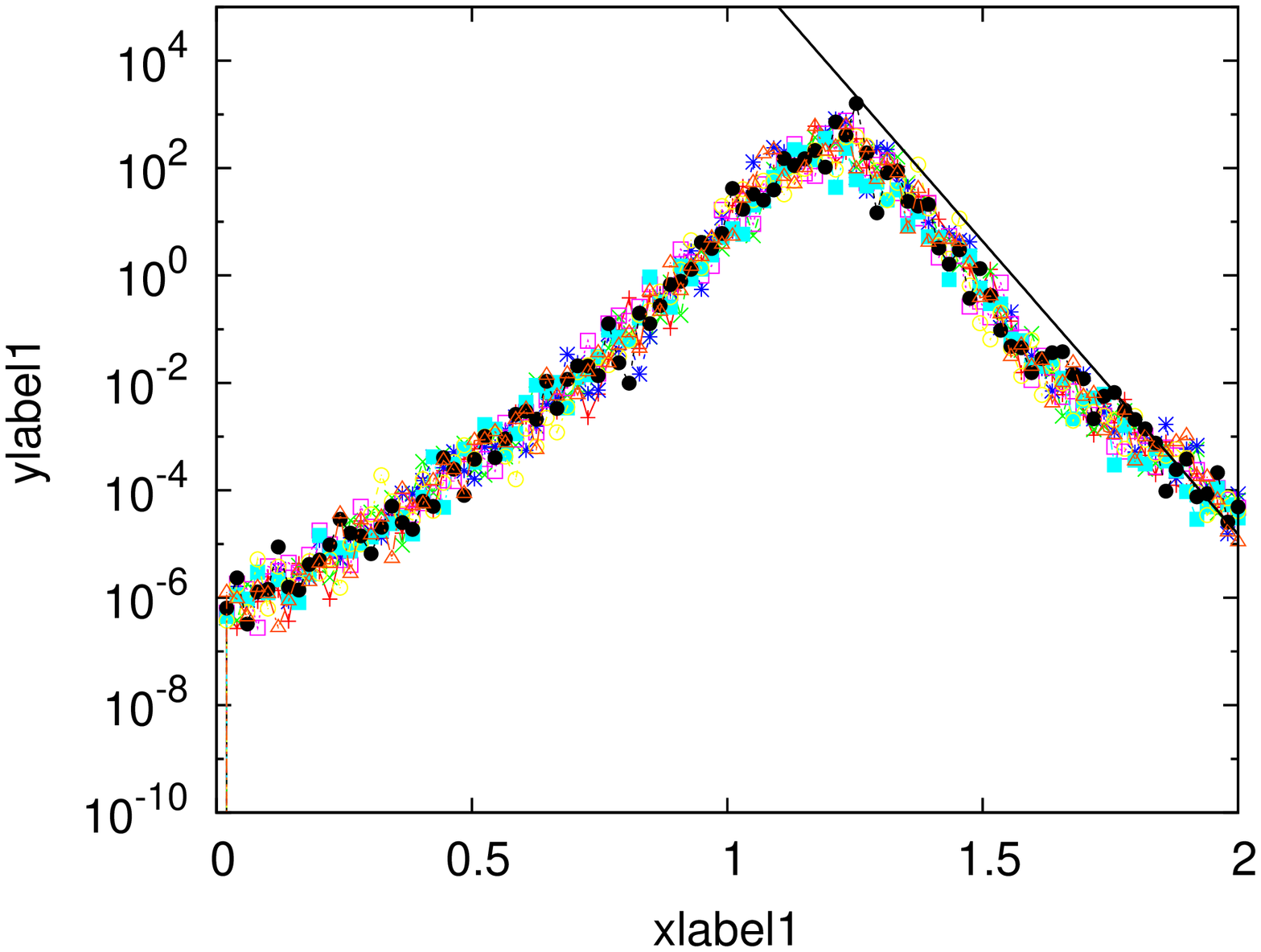}
\includegraphics[height=0.3\textwidth,angle=270]{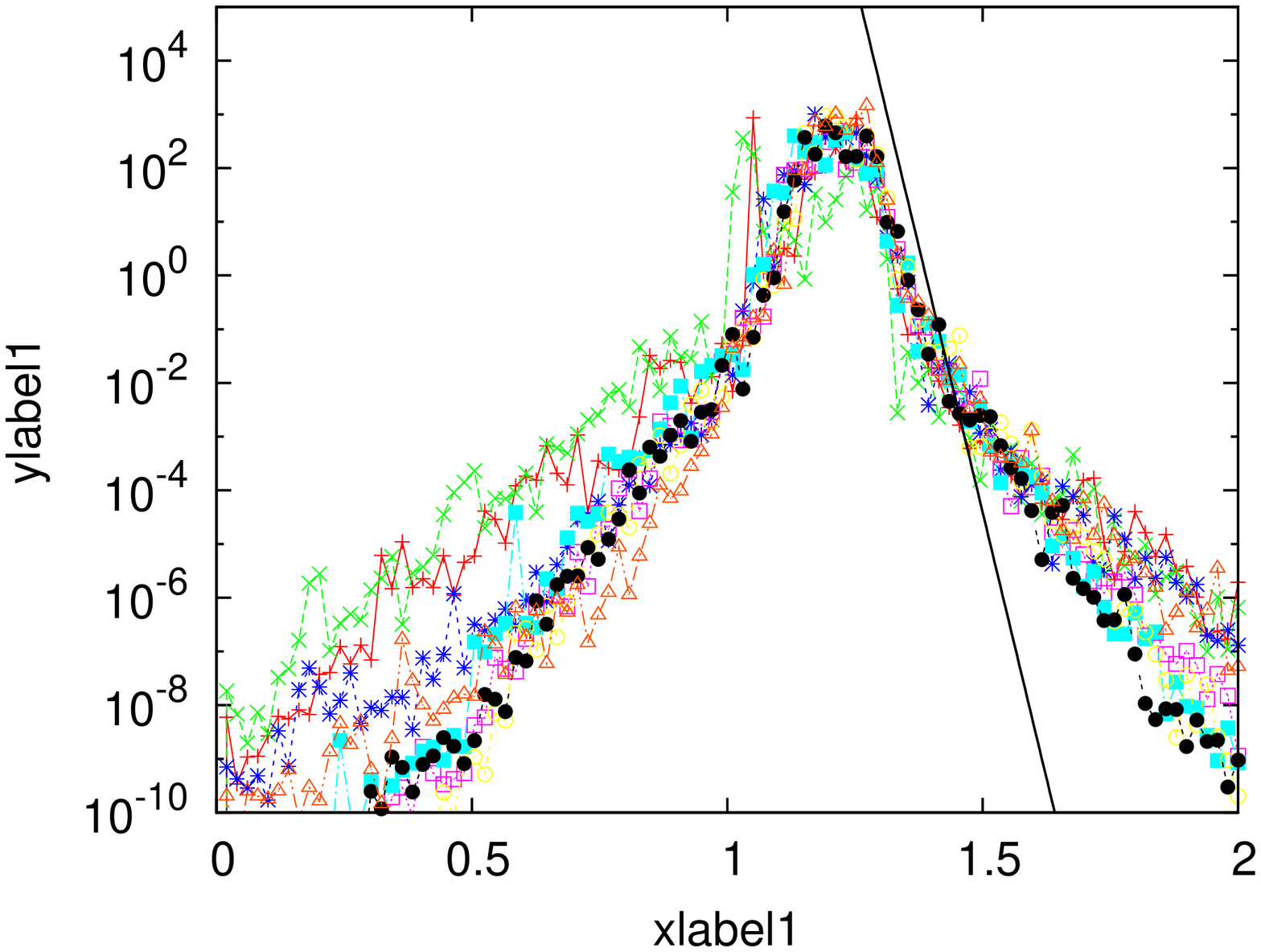}
\caption{Diffusion $D_{zk}(\omega)$ in $z$-variable for particles $k=1,...,8$,
shown by different color symbols,
in model C1 at $L=8$ and  $\gamma=1.0965\cdot10^{-2}$ (left), 
$1.5849\cdot10^{-3}$ (middle), $1.2023\cdot10^{-4}$ (right).
The straight line shows the dependence 
$D_{zk} \propto \exp(-\omega/\sqrt{\gamma})$. }
\label{fig:difz}
\end{figure}

An indirect support to the view point according to which
at $t_{max}=10^8$ we still did not reach the asymptotic 
spreading exponent $\alpha=0.3636$ is based on the 
numerical computation of the diffusion rate 
in an additional effective degree of freedom
described by the equation $d z_k/dt = q_k \sin (\omega t)$,
where $q_k$ are dynamical variables in model C1 (\ref{eq:modc}).
Solving these equations in parallel with the dynamical
equations of motion for $q_k$ we determine the effective diffusion constant
$D_{zk}(\omega)$ for each particle $k$ at $L=8$. 
To suppress regular quasiperiodic
oscillations we use the window averaging method described in
\cite{Chirikov_etal-85} computing first
the average 
$\bar{z}_k(j)=\int_{jT}^{(j+1)T} z_k(t) \sin^6(2\pi t/T) d t/ 
\int_{jT}^{(j+1)T} \sin^6(2\pi t/T) d t$ over time interval $T=10^6$
and determining the diffusion for each $k$ via the relation
$D_{zk}(\omega) = \sum_{j'>j \geq 1} (\bar{z}_k(j) - \bar{z}_k(j'))^2/((j'-j)T)$.
The computation is done for one trajectory 
with total time $t=10^7$. The initial particle energies 
are chosen to be $E_k=p_k^2/2+q_k^4/4 \approx 1$
at $q_k=0$.
At $\gamma=0$ we have the particle action
$I_k=\Gamma(1/4) E_k^{3/4}/(2 \sqrt{\pi} \Gamma(7/))$
and nonlinear frequency 
$\omega_k=\partial H/\partial I_k = 
8\sqrt{\pi}\Gamma(1/4)E_k^{1/4}/(3 \Gamma(7/8)) 
\approx 1.2 E_K^{1/4} \approx 1.2$. The dependence of
$D_{zk}$ on frequency $\omega$ is shown in Fig.~\ref{fig:difz}
for all $k$ values. In fact, $D_{zk}(\omega)$
gives us the spectral density of an effective noise produced by
dynamical chaos. According to the results obtained for the
modulational diffusion \cite{Chirikov_etal-85},
the spectrum of $D_{zk}(\omega)$ is expected to
have  a plateau of width
$\Delta \omega$ centered at the resonance $\omega_k \approx 1.2$,
followed by an exponential drop 
$D_{zk}(\omega) \propto 
(1/\Delta \omega) \exp(-|\omega -\omega_k|/\Delta \omega)$.
In the picture of triplet resonance we have 
$\Delta \omega \sim \sqrt{\gamma}$.
The data of Fig.~\ref{fig:difz}
are in a satisfactory agreement with such
a picture showing a decrease of the plateau size 
with the decrease of $\gamma$. The plateau is
followed by an exponential drop.
However, at $\gamma \sim 0.01$ the spectral
width $\Delta \omega $ is still rather large
being comparable with $\omega_k \sim 1$.
At such spectral width even the oscillators 
that are not directly involved in the triplet resonance
still will be affected by it. This is probably the reason
why up to $\gamma \sim 0.02$ we have 
the spreading of chaos with the exponent $\alpha \approx 0.6$
corresponding to a usual diffusion $D \propto \gamma^2$
in model C2. At $\gamma \sim 10^{-4}$ the spectral width
becomes notably smaller than unity
but one needs to go to enormously large times $t_{max} \sim 10^{26}$
to reach such effective values of $\gamma$
during spreading of chaos. The value $K \sim \gamma \sim 10^{-5}$
where there is a change in the dependence $D(K)$ detected
by Chirikov and Vecheslavov (see Fig.~1 in \cite{Chirikov-Vecheslavov-97})
would require times at least as large as $t_{max} \sim 10^{33}$.
Definitely such times remain out of reach of modern
computations.

On the basis of presented results and discussions
we can say that the spreading of chaos over the nonlinear 
oscillator lattice of model C2  (\ref{eq:modc}) goes in a subdiffusive way
(\ref{eq:spread}) with the exponent $\alpha \approx 0.55$ up to times
$t \sim 10^8$. In view of the result of Chirikov and Vecheslavov
for the fast Arnold diffusion (\ref{eq:difdk}) 
\cite{Chirikov-Vecheslavov-97} it is possible that the 
exponent will go down to $\alpha \approx 0.36$ at times
$t > 10^{26}$. The properties of chaos spreading
behind times $10^{33}$ remain absolutely unknown.
During this anomalous slow growth of the wave packet size,
the chaotic spreading 
follows the Arnold web of tiny chaotic layers
propagating mainly along mostly thick ones.
However, from time to time a trajectory can go inside
thinner layers that leads to a strong drop of local diffusion 
and propagation rates, as well as a significant drop of LE
(see, e.g., Fig.~\ref{fig:localle}). It is quite possible that
in this regime the energy $E_k$ distribution over the populated 
modes $N(t)$ is still more or less homogeneous, 
as it is seen in Fig.~\ref{fig:spread}, however,
we expect this state to be not ergodic within these $N$ modes since
chaos is presumably confined inside some ``porous medium''
of Arnold web with very complex structure and topology. 
In course of spreading, the energy per excited oscillator 
goes down to zero, so  that such a process can be considered as
an unusual non-ergodic cooling.

\section{Slow diffusion in Hamiltonian systems as  deterministic rheology}
\label{sec:rheology}

The spreading of chaos discussed above goes in a very slow way.
In this section we explore a parallel with slow rheology processes 
characterized
by small values of the Deborah number \cite{Reiner-64}
\begin{equation}
D_R = t_r/t_{obs} \sim 1/(\lambda t_{obs}) \ll 1 \;\;, 
\label{eq:deborah}
\end{equation}
where $t_r$ is a time scale of local relaxation process
and $t_{obs}$ is a time of observation.
The values of $D_R \ll 1$ correspond to a liquid-like phase
while $D_R \gg 1$ appears for the solid phase.
At our initial state with one or few excited oscillators
we have the relaxation time to be comparable with the inverse LE
$t_r \sim 1/\lambda \sim 1$, while the observation
time of spreading is $t_{obs} \sim 10^8$ for our numerical
simulations. Thus we have extremely small values of $D_R \sim 10^{-8}$
for our studies. The parallels with rheology
processes, which are actively studied in soft matter
and porous materials (see e.g \cite{Malkin-06,Rao-07}), 
can be build on the basis of the following arguments:
a)in rheology the flow processes are characterized by 
small $D_R$ values that is exactly the  case for chaos spreading in model C2;
b)often a spreading in a porous media is described
by a nonlinear diffusion for a density $\rho(x,t)$ \cite{Barenblatt-03}:
\begin{equation}
\partial \rho/\partial t = D_0 
\partial (\rho^a \partial \rho/\partial x)/\partial x 
\label{eq:nldifeq}
\end{equation}
and it was shown recently that this equation gives 
a good phenomenological description of chaos spreading in nonlinear lattices;
\cite{Mulansky-Pikovsky-10};
c)the Arnold web of chaotic resonance layers forms some kind of a porous
media along which energy can spreads to larger and larger
sizes. Recent experiments on gel formed by  attractive colloidal 
hard spheres, suspended in an aqueous solvent, 
show that the spreading of gel is indeed well described by 
such type of a nonlinear diffusion equation (\ref{eq:nldifeq})
with a nonlinear flux term \cite{Cipelletti-11}. The theoretical models
of rheology flow try to explain such a spreading by
phenomenological statistical models with disorder and metastability
(see e.g. \cite{Sollich-97,Sollich-06}).
In contrast to such statistical models, our
``rheology'' of chaos spreading has purely dynamical
and deterministic  origin.

The value of $D_R$ given above should be considered as
a global simplified estimate.
It is also important to see how $D_R$ varies with time $t_{obs}$ of 
spreading duration.
For the model C2 we have 
$\lambda \sim I_k \sim  E_k^{3/4} \sim N^{-3/4} \sim t_{obs}^{-3 \alpha/8}$
and hence from (\ref{eq:deborah}) we find 
$D_R(t) \sim 1/t_{obs}^{1-3 \alpha/8} \propto 1/t_{obs}^{0.78} \gg 1$.
Thus in this model $D_R \rightarrow 0$ at $t_{obs} \rightarrow \infty$
that argues in a favor of continuation of spreading at infinitely large times.
The same criterion applied to the DANSE model, which describes
the Anderson model with nonlinearity $\beta |\psi|^2$ and was
studied in \cite{Pikovsky-Shepelyansky-08},
gives $\lambda \sim I \sim \beta/N \sim \beta/t_{obs}^{\alpha/2}$
and thus 
$D_R \sim 1/ (\beta t_{obs}^{(1-\alpha/2)})$ still goes to zero
in the limit of large times ($\alpha \approx 1/3$ for DANSE).
The above arguments show that for the nonlinearity
$\beta |\psi|^{2a}$ studied in \cite{Mulansky-Pikovsky-10} 
we have $\lambda \sim I^a \sim 1/t_{obs}^{\alpha a/2}$
and $D_R \sim 1/t_{ons}^{(1-\alpha a /2)} \rightarrow 0$
even for $a=2,3$ (see corresponding values of $\alpha$
given in \cite{Mulansky-Pikovsky-10}). Indeed, the numerical results 
of \cite{Mulansky-Pikovsky-10} show an infinite spreading
for such values of $a$.

The above discussion shows that weakly nonintegrable many-body Hamiltonian
systems give a new interesting example of 
rheology of chaotic dynamics. These systems are ruled
by purely deterministic and rather simple Hamiltonian equations of motion.
Exploring further statistical properties of such a  
deterministic rheology, generated by 
Hamiltonian many-body dynamics, is 
an important task for future studies.

\section{Conclusion}

In this paper we studied properties of non-integrable Hamiltonian lattices 
focusing on the regimes of very weak non-integrability. Our main results 
are scaling relations for the probability to observe strong chaos 
with  the largest possible Lyapunov exponent. 
This probability is proportional to 
the product of the coupling parameter and the lattice length, 
while the Lyapunov exponent scales as a square root of the coupling constant. 
This behavior is explained by the observation that strong chaos is mainly 
due to resonances that appear when three neighboring sites occasionally 
have close frequencies. Because both the frequency mismatch and 
the characteristic time scale of the resonance are proportional 
to a square root of the perturbation parameter, the relations 
above directly follow from this scaling. 

Furthermore, we confirm previous calculations 
showing that the diffusion time scale at weak non-integrability
is much larger than the inverse Lyapunov exponent, and 
relate this to a weak diffusion inside the weak chaos component.
The measure of this component decreases only algebraically
with the strength of nonlinear coupling between nonlinear oscillators.
The obtained results are in a good agreement with the fundamental
finding of Chirikov and Vecheslavov 
\cite{Chirikov-Vecheslavov-90,Chirikov-Vecheslavov-93,Chirikov-Vecheslavov-97}
who first discovered this regime,
with only algebraic decrease of the measure of chaos and
diffusion rate at rather small perturbations, and named it
the fast Arnold diffusion.

We also studied the spreading of chaos in such coupled nonlinear lattices 
showing that the spreading goes in an anomalous subdiffusive way.
The link between the exponent of this spreading and
the fast Arnold diffusion are also determined.

As already mentioned in the introduction, one has to 
distinguish weakly nonlinear and weakly non-integrable systems. 
There is, however, some analogy between the dynamics of 
weakly nonintegrable lattices studied in this paper and 
random lattices with weak nonlinearity~\cite{Basko-10,Pikovsky-Fishman-10,%
Johansson-Kopidakis-Aubry-10,Pikovsky-Shepelyansky-08,%
Garcia-Mata-Shepelyansky-09}. 
We consider homogeneous lattices, where resonances appear randomly 
due to random choice of initial conditions. 
In random weakly nonlinear lattices resonances are 
determined by a lattice disorder. 
So in both cases one can expect that chaos is 
mainly sitting on resonances. For nonlinear homogeneous lattices, 
resonances can ``move'' as the energies on different lattice sites 
vary, while in weakly nonlinear disordered lattices the resonances 
are due to disorder and thus are ``pinned''. The properties
of chaos spreading in the latter case require separate investigations.

\begin{acknowledgements}
We thank S. Fishman for useful discussions.
A.P. thanks UPS, Toulouse for hospitality and support,
DLS thanks Univ. of Potsdam for hospitality during visits in 2009, 2010. 
The work was supported by DFG via grant PI220/12. 
We thank ZEIK (Univ. Potsdam) and HLRS Stuttgart 
for providing the computer facilities.
\end{acknowledgements}


\end{document}